\documentclass[twocolumn,trackchanges]{aastex631}
\usepackage{bm}
\usepackage{gensymb}
\usepackage{tabularx}

\begin{document}

%\title{Template \aastex Article with Examples: 
%v6.3.1\footnote{Released on March, 1st, 2021}}
%\title{Dips and Coronal Evolution During the 2025 Failed Outburst of IGR~J17091–3624}
\title{The 2025 Failed Outburst of IGR~J17091–3624: Spectral Evolution and the Role of Ionized Absorbers}

\author[0000-0002-5966-4210]{Oluwashina K. Adegoke}
\affiliation{Cahill Center for Astronomy \& Astrophysics, California Institute of Technology,
Pasadena, CA 91125, USA}

\author[0000-0003-3828-2448]{Javier A. Garc\'ia}
\affiliation{X-ray Astrophysics Laboratory, NASA Goddard Space Flight Center, Greenbelt, MD 20771, USA}
\affiliation{Cahill Center for Astronomy \& Astrophysics, California Institute of Technology,
Pasadena, CA 91125, USA}

\author[0000-0003-4216-7936]{Guglielmo Mastroserio}
\affiliation{Dipartimento di Fisica, Universit\`a Degli Studi di Milano, Via Celoria, 16, Milano, 20133, Italy}
\affiliation{Scuola Universitaria Superiore IUSS Pavia, Palazzo del Broletto, piazza della Vittoria 15, I-27100 Pavia, Italy}

\author[0000-0002-0273-218X]{Elias Kammoun}
\affiliation{Cahill Center for Astronomy \& Astrophysics, California Institute of Technology,
Pasadena, CA 91125, USA}

\author[0000-0002-8908-759X]{Riley M. T. Connors}
\affiliation{Department of Physics, Villanova University, 800 E. Lancaster Avenue, Villanova, PA 19085, USA}

\author[0000-0002-5872-6061]{James F. Steiner}
\affiliation{Center for Astrophysics \textbar\ Harvard \& Smithsonian, 60 Garden Street, Cambridge, MA 02138, USA}

\author[0000-0003-2992-8024]{Fiona A. Harrison}
\affiliation{Cahill Center for Astronomy \& Astrophysics, California Institute of Technology,
Pasadena, CA 91125, USA}

\author[0000-0002-5341-6929]{Douglas J. K. Buisson}
\affiliation{Independent}

\author[0000-0001-7532-8359]{Joel B. coley}
\affiliation{Department of Physics and Astronomy, Howard University, Washington, DC 20059, USA}
\affiliation{CRESST and NASA Goddard Space Flight Center, Astrophysics Science Division, 8800 Greenbelt Road, Greenbelt, MD, USA}

\author[0000-0003-0870-6465]{Benjamin M. Coughenour}
\affiliation{Department of Physics, Utah Valley University, 800 W. University Parkway, MS 179, Orem UT 84058, USA}

\author[0000-0003-4583-9048]{Thomas Dauser}
\affiliation{Dr.\ Remeis-Sternwarte~\&~ECAP, Universit\"at Erlangen-N\"urnberg, Sternwartstr. 7, 96049 Bamberg, Germany}

\author[0000-0001-9349-8271]{Melissa Ewing}
\affiliation{School of Mathematics, Statistics, and Physics, Newcastle University, Newcastle upon Tyne NE1 7RU, UK}

\author[0000-0002-5311-9078]{Adam Ingram}
\affiliation{School of Mathematics, Statistics, and Physics, Newcastle University, Newcastle upon Tyne NE1 7RU, UK}

\author[0000-0003-0172-0854]{Erin Kara}
\affiliation{MIT Kavli Institute for Astrophysics and Space Research, MIT, 70 Vassar Street, Cambridge, MA 02139, USA}

\author[0000-0002-9633-9193]{Edward Nathan}
\affiliation{NASA Postdoctoral Program Fellow, NASA Goddard Space Flight Center, Code 662, Greenbelt, MD 20771, USA}

\author[0009-0003-8610-853X]{Maxime Parra}
\affiliation{Department of Physics, Ehime University, 2-5, Bunkyocho, Matsuyama, Ehime 790-8577, Japan}

\author[0000-0003-2686-9241]{Daniel Stern}
\affiliation{Jet Propulsion Laboratory, California Institute of Technology, Pasadena, CA 91109, USA}

\author[0000-0001-5506-9855]{John A. Tomsick}
\affiliation{Space Sciences Laboratory, 7 Gauss Way, University of California, Berkeley, CA 94720-7450, USA}

\begin{abstract}
% JG version:
IGR~J17091–3624 is the only black hole X-ray binary candidate, aside from the well-studied black hole system GRS~1915+105, observed to exhibit a wide range of structured variability patterns in its light curves. In 2025, the source underwent a “failed” outburst: it brightened in the hard state but did not transition to the soft state before returning to quiescence within a few weeks. During this period, IGR~J17091–3624 was observed by multiple ground- and space-based facilities. Here, we present results from six pointed NuSTAR observations obtained during the outburst. None of the NuSTAR light curves showed the exotic variability classes typical of the soft state in this source; however, we detected, for the first time, strong dips in the count rate during one epoch, with a total duration of $\sim4\,\mathrm{ks}$ as seen by NuSTAR. Through spectral and timing analysis of all six epochs, we investigate the hard-state spectral evolution and the nature of the dips. A clear evolution of the coronal properties with luminosity is observed over all six epochs, with clear signatures of relativistic disk reflection which remain largely unchanged across the first five epochs. The first five epochs also show a strong and stable quasi-periodic oscillation (QPO) feature in the power spectra. The dips observed in Epoch 5 are consistent with partial obscuration by ionized material with a column density $N_{\mathrm{H}} \approx 2.0 \times 10^{23}\,\mathrm{cm^{-2}}$. We discuss possible origins for this material and place constraints on the orbital parameters and distance of the system.
\end{abstract}



\section{Introduction} \label{sec:one}
At the onset of an outburst, a black hole X-ray binary (BHXB) typically rises in the \textit{hard state} where emission from the corona dominates, believed to be produced by inverse Compton scattering of lower energy disk photons \citep{1976MNRAS.175..613S, 1973blho.conf..343N} or internal synchrotron photons \citep{2009ApJ...690L..97P, 2009MNRAS.392..570M, 2011ApJ...737L..17V}. It then transitions to the \textit{soft state}---dominated by low-energy disk blackbody photons---passing through an \textit{intermediate state} before returning to quiescence through the intermediate and hard states, respectively \citep[e.g.,][]{2007A&ARv..15....1D, 2011BASI...39..409B, 2022hxga.book....9K}. While an outburst can last several months, time spent in the intermediate state is typically significantly shorter than in the hard and soft states. BHXBs are sometimes known to go through so-called ``failed'' outbursts. During a failed outburst, the source rises in the hard state but does not transition to the soft state before returning to quiescence \citep[see e.g.,][]{2019ApJ...885...48G, 2021MNRAS.507.5507A}. 

IGR J17091-3624 was discovered in 2003, although archival data showed that the source has had a number of previous outbursts, starting from 1994 \citep{2003ATel..149....1K, 2003ATel..150....1R, 2003ATel..160....1I, 2016ApJS..222...15T}. Over the past three decades, it has undergone close to a dozen outbursts \citep[see e.g.,][]{2003ATel..149....1K, 2009ApJ...690.1621C, 2011ApJ...742L..17A, 2016ApJS..222...15T}. Since the launch of NuSTAR \citep{2013ApJ...770..103H}, two such events have been observed—one in 2016 and another in 2022 \citep{2017ApJ...851..103X, 2024ApJ...963...14W}. The most recent outburst, in 2025, appears to be the first reported failed outburst from the source. While the spectro-temporal behavior of IGR~J17091-3624 is fully consistent with a BHXB, and rather unlikely for the system to host a neutron star, it is not a dynamically-confirmed black hole and is thus a black hole candidate. IGR~J17091-3624 is peculiar, as it is the only BHXB candidate observed to show structured variability patterns in its light curves besides the BHXB GRS~1915+105. While for GRS 1915+105 the light curve variability has been grouped into at least fourteen classes, the light curve variability in IGR J17091-3624 is grouped into ten so far. Of the ten variability classes, seven resemble those from GRS 1915+105 including the famous ``heartbeat'' variability class---class IV and class $\rho$ in IGR~J17091-3624 and GRS~1915+105, respectively---mimicking an electrocardiogram \citep[e.g.,][]{2000A&A...355..271B, 2002MNRAS.331..745K, 2005A&A...435..995H, 2017MNRAS.468.4748C, 2018MNRAS.476.1581A, 2020MNRAS.492.4033A, 2024ApJ...963...14W}. High-frequency quasi-periodic oscillations (QPOs) are detected at similar frequencies, $\sim66\,\mathrm{Hz}$, in both sources although the variability in IGR~17091-3624 is generally faster than in the corresponding GRS~1915+105 class \citep{2012ApJ...747L...4A, 2017MNRAS.468.4748C}. Like in GRS~1915+105, spectral absorption lines from highly ionized iron have been detected in IGR~J17091-3624, indicating the possible presence of an outflowing disk wind near the black hole \citep{2012ApJ...746L..20K, 2024ApJ...963...14W}. 
%In its bright 2011 outburst, Chandra's High Energy Transmission  Grating (HETG) spectrum revealed, while in the intermediate state, the presence of an absorption line in the source at $6.91\pm{0.01}\,\mathrm{keV}$, implying an extreme outflow of $\sim0.03\,c$ if associated with blueshifted Fe \textsc{xxv} line \citep{2012ApJ...746L..20K}. 
IGR~J17091-3624 is of particular interest because it tends to bridge the accretion flow properties of a peculiar source like GRS~1915+105 to those from more ``normal'' BHXBs---going through outburst cycles with evolutionary patterns typical of BHXBs. In principle, it may hold the clue to understanding the origin of these exotic variability behavior as compared to the behavior of standard BHXBs. 

Limit-cycle or radiation pressure instability in the inner accretion disk, when a source is accreting at a significant fraction of its Eddington luminosity, is commonly thought to be responsible for the structured variability patterns \citep[e.g.,][]{2000ApJ...535..798N, 2004MNRAS.349..393D, 2011ApJ...737...69N}. This model seems to work well for GRS~1915+105 as it can sometime attain super-Eddington luminosities. Because IGR~J17091-3624 is about a factor of $20-30$ fainter (in its peak flux) than GRS~1915+105 and it shows the kind of exotic variability patterns known with GRS~1915+105, the high accretion rate criteria for the structured light-curve variability patterns has been questioned. The mass of IGR~J17091-3624 is not known, also, the distance and orbital parameters of the system are not reliably constrained. Thus, a significantly high accretion rate scenario could imply that IGR~J17091-3624 either harbors one of the least massive black holes known ($<3\,M_{\odot}$---for a distance less than $17\,\mathrm{kpc}$) or it is much further away---up to $\sim23\,\mathrm{kpc}$ in the Galactic disk or so distant that it lies outside of the galaxy \citep{2011ApJ...742L..17A, 2024ApJ...963...14W}. These two scenarios are not in agreement with the constraints obtained from a number of model-dependent analyses \citep[see e.g.,][]{2011A&A...533L...4R, 2012MNRAS.422L..91W}. 
%For example, while \citet{2011A&A...533L...4R} inferred a distance of $\sim11-17\,\mathrm{kpc}$ to the system, \citet{2015ApJ...807..108I}, by probing the evolution of its photon index, QPO and broadband spectra during its 2011 outburst, estimated the black hole mass in IGR~J17091-3624 to be in the range $8.7-15.6\,M_{\odot}$.

The onset of a new, but short-lived, outburst was detected in IGR~J17091-3624 around February of 2025 \citep{2025ATel17034....1R} and was monitored by several space missions including NuSTAR and NICER \citep{2016SPIE.9905E..1HG}. NuSTAR observed the source six times, all in the hard state, during the period between February and April. Using data from an IXPE \citep{2022JATIS...8b6002W} observation during the outburst, \citet{2025MNRAS.tmp..846E} measured, in the $2-8\,\mathrm{keV}$ band, a polarization degree of $9.1\pm{1.6}\%$ and a polarization angle of $83\pm{5}\degree$ for the source in the hard state. The high polarization degree was attributed to either the system being highly inclined, bulk motion of the electrons in the corona and/or scattering in an optically thick disk wind.
\begin{figure}%[ht!]
\includegraphics[width=0.40\textwidth, angle=0, trim={0.0cm 0.0cm 5cm 1cm}]{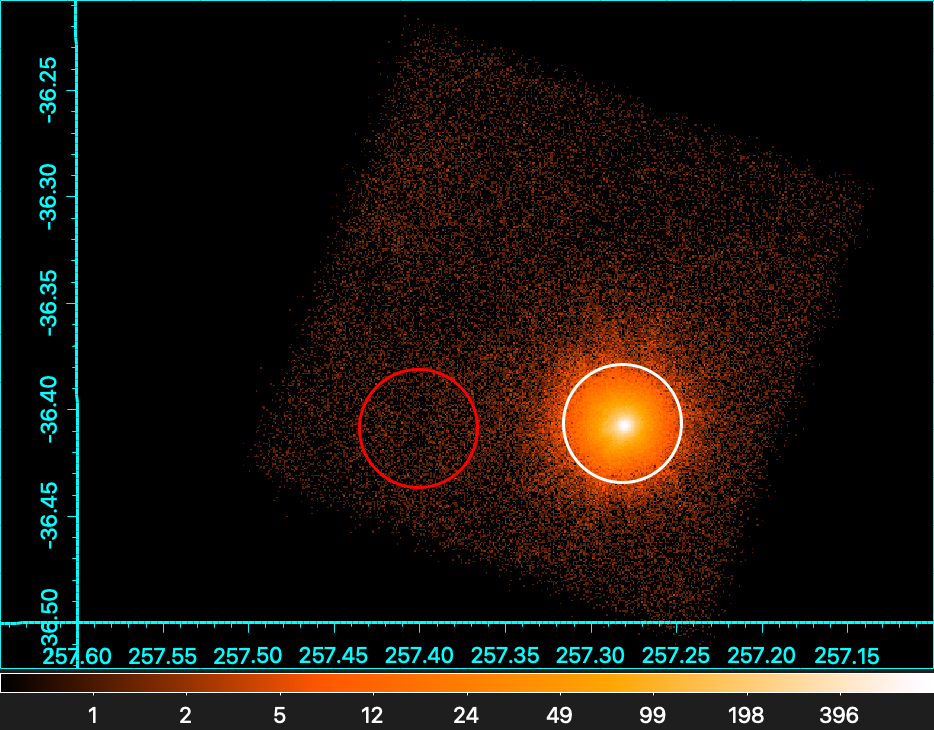}
\caption{Exposure-corrected NuSTAR image of IGR~J17091-3624 from Epoch 4 (The color map is logarithmically scaled). The source and background regions, each extracted from a 100'' radius, are captured by the white and red circles, respectively. The color scale of the image is in counts per pixel. The horizontal and vertical axes are the J2000 coordinates of the system.} 
\label{fig:nustar_image_ep4}
\end{figure}
 In this paper, we probe the spectral evolution of the source during its 2025 outburst,
 %in relation to its previously reported outbursts,
 using exclusively the NuSTAR data. We further probe the possible origin of the recurrent light-curve dips seen in one of the epochs. While the source was significantly monitored by NICER during the outburst, several of the observations were carried out during ``orbit day'' and at a period when the NICER measurement/power units (MPUs) were being reconfigured after the light-leak repair on the telescope\footnote{\url{https://www.nasa.gov/missions/station/nicer-status-updates/}}. Analysis of these data requires special care and is therefore deferred to a future publication. The paper is structured as follows. In Section 2, we describe the observation and data reduction procedure. In Section 3, we present the data analysis and the results. In Section 4, we discuss the results and their implications and in Section 5, we summarize our main findings.
 
%%%
\begin{table}%[H]
\caption{Log of the NuSTAR observations of \rm{IGR~J17091-3624} used.} % title of table
\centering % used for centering table
%\resizebox{\columnwidth}{!}{%
\begin{tabular}{l l l l } % %centered columns(2)
\hline\hline %inserts double horizontal lines
\textbf{Epoch} & \textbf{ObsID} & $\bm{T_{ \rm start} }$ & $\bm{T_{\rm exp}}$\\ [0.5ex] %  & $\bm{R_{\rm src}}$ 
 & & \textbf{(UTC)} & \textbf{(ks)}\\ [0.5ex]% & $\bm{(\arcsec)} $
%heading
\hline\hline % inserts single horizontal line
1 & 81002342002 & 2025-02-16 13:41:05 & 33.5\\% & 100
%& \textit{NICER}& 3133010104 & 2021-01-23 17:26:25 & 1.2\\
%& \textit{NICER}& 3133010105 & 2021-01-23 23:39:07 & 8.6\\
2 & 81002342004 & 2025-02-18 19:51:10 & 33.6\\% & 150
%& \textit{NICER}& 3133010116 & 2021-02-05 04:51:20 & 5.3\\
3 & 81002342006 & 2025-02-20 10:16:11 & 32.7\\% & 100
%& \textit{NICER}& 3558010501 & 2021-02-20 03:52:17 & 1.5\\
4 & 81002342008 & 2025-03-07 14:01:06 & 21.0\\% & 100
%& \textit{NICER}& 3558010902 & 2021-03-07 04:11:11 & 4.5\\
5 & 81002342010 & 2025-03-08 23:31:12 & 18.9\\% & 100
%& \textit{NICER}& 4133010103 & 2021-03-26 03:22:39 & 6.1\\
6 & 91102310002 & 2025-04-20 02:46:11 & 28.3\\% & 100
%& \textit{NICER}& 4133010109 & 2021-04-01 00:21:20 & 16.1\\
\hline 
\hline %inserts single line

\end{tabular}
%}
\label{tab:log} % is used to refer this table in the text
\end{table}
%%%
%\section{Manuscript styles} \label{sec:style}
\section{Observations and Data Reduction} \label{sec:two}
%\subsection{NuSTAR} \label{subsec:two-one}
We analyze the six NuSTAR observations taken between February 16 and April 20, 2025 (see Table \ref{tab:log}).
%%%%
\begin{figure*}%[ht!]
\includegraphics[width=0.9\textwidth, angle=0, trim={1.0cm 0.5cm 3.0cm 0.5cm}]{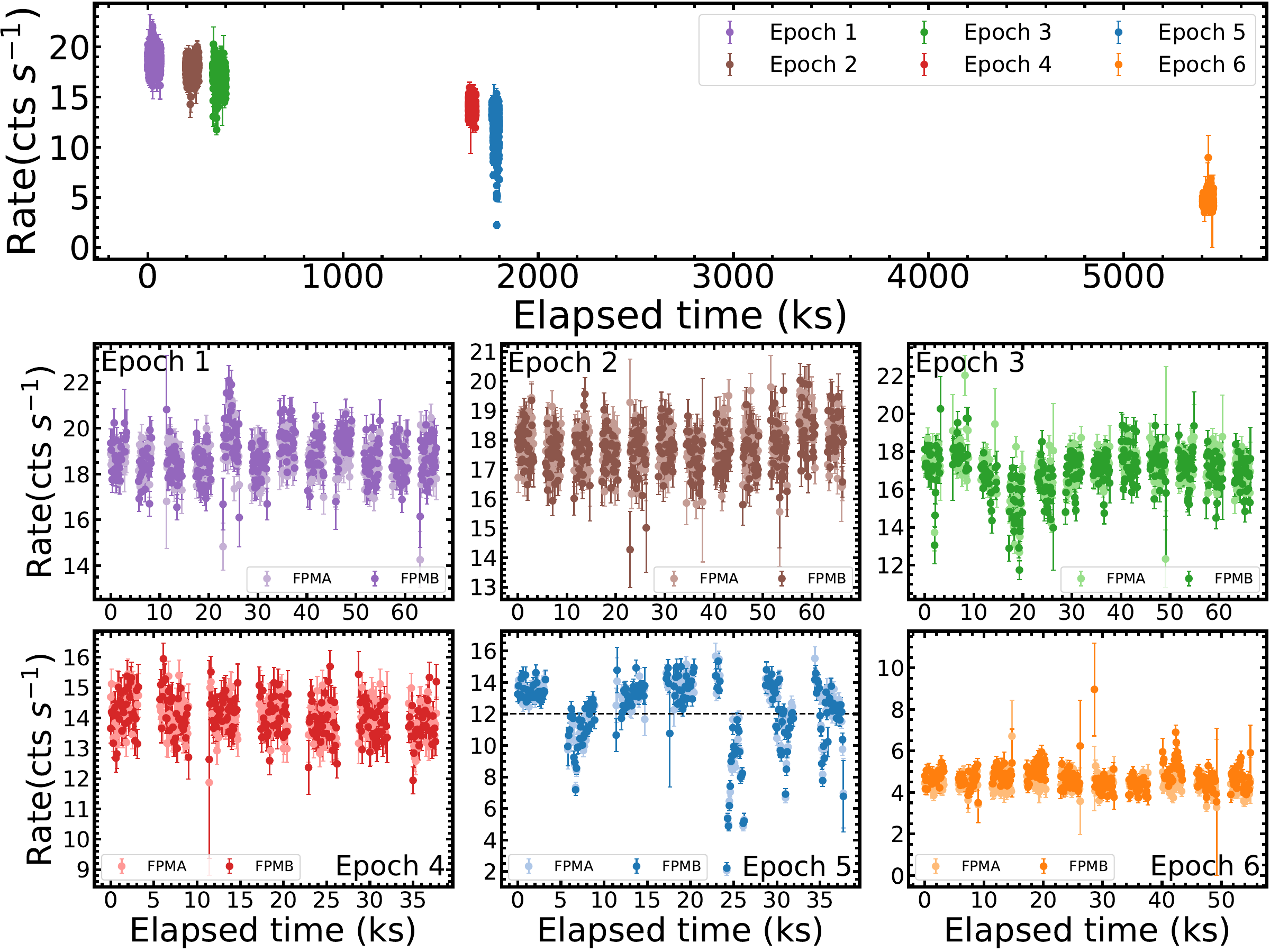}
\caption{NuSTAR FPMB light curves for all six epochs combined (topmost panel) and plotted separately (lower panels). All the light curves have been binned to $100\,\mathrm{s}$. For Epoch 5, the dashed horizontal line separates the persistent interval from the dip interval.} 
\label{fig:lc_nustar}
\end{figure*}
%%
%\begin{figure*}
%%\vspace{-10pt}
%\includegraphics[width=1.0\textwidth, angle=0, trim={2cm 1cm 2cm 1cm}]{nu_epoch_1-6_lc_igr.pdf}\\
%\includegraphics[width=.3\textwidth, angle=0, trim={1.0cm 0cm 1.5cm 0.5cm}]{nu-epoch1_lc_igr.pdf}
%\includegraphics[width=.3\textwidth, angle=0, trim={1.0cm 0cm 1.5cm 0.5cm}]{nu-epoch2_lc_igr.pdf}
%\includegraphics[width=.3\textwidth, angle=0, trim={1.0cm 0cm 1.5cm 0.5cm}]{nu-epoch3_lc_igr.pdf}\\
%%\hspace{-10pt}
%\includegraphics[width=.3\textwidth, angle=0, trim={1.0cm 0cm 1.5cm 0.5cm}]{nu-epoch4_lc_igr.pdf}
%\includegraphics[width=.3\textwidth, angle=0, trim={1.0cm 0cm 1.5cm 0.5cm}]{nu-epoch5_lc_igr.pdf}
%\includegraphics[width=.3\textwidth, angle=0, trim={1.0cm 0cm 1.5cm 0.5cm}]{nu-epoch6_lc_igr.pdf}\\
%\caption{NuSTAR light curves for all six epochs. All the light curves have been binned to $100\,\mathrm{s}$. For Epoch 5, the dashed horizontal line separates the persistent interval from the dip interval.}
%%N.B: For epoch 3, the \textit{NICER} data are from two non-continuous ObsIDs.}
%\label{fig:lc_nustar}
%\end{figure*}
Data reduction was carried out using the standard pipeline Data Analysis Software, \textsc{nustardas} v.2.1.4a and \textsc{caldb} v20250317. Event files and images were generated with the \texttt{nupipeline} command. 
Source products were extracted from a circular region of radius 150'' centered on the source for Epoch 2. For Epochs 1 and 3, a smaller extraction region of radius 100'' was used because of the presence of stray light contamination. For Epochs 4-6, the source flux has dropped so the extraction region radius was also chosen to be 100''. In all cases, the background was extracted from a source-free region of the same size as the source. Although the overall effect of the stray light contamination in the data of Epochs 1 and 3 is minimal, the background products for these two epochs were extracted from source-free stray-light contaminated regions to further mitigate the effect. Figure \ref{fig:nustar_image_ep4} shows the exposure-corrected NuSTAR image of Epoch 4.
%depending on the brightness of the source during that epoch. 
 %For ObsIDs with incident count rate in excess of $100\mathrm{cts\,s^{-1}}$, we included the keyword `statusexpr="(STATUS==b0000xxx00xxxx000)\&\&(SHIELD ==0)"'  in the \texttt{nupipeline} command run as suggested in the ``NuSTAR Frequently Asked Questions'' page\footnote{\url{https://heasarc.gsfc.nasa.gov/docs/nustar/nustar_faq.html}}. This is to prevent good source count from being vetoed by NuSTAR's noise filter. 
The \texttt{nuproducts} task was subsequently employed to generate source and background spectra, and light curves as well as instrumental responses. Because the source count-rate is low, there is no noticeable difference between FPMA and FPMB spectra at low energies; thus multi-layer insulation (MLI) correction was not applied during spectral fitting \citep[see e.g.,][]{2020arXiv200500569M}.

%\subsection{NICER} \label{subsec:two-two}
%The 2021 outburst of GX~339-4 was extensively monitored by NICER---with 262 individual pointings between January and November of that year. A number of these observations were simultaneous or quasi-simultaneous with the 13 targeted NuSTAR observations of the source over the duration of the outburst (see Table \ref{tab:log}).

%The data reduction followed standard procedure, using \textsc{nicerdas} version 10a, as outlined in the NICER Data Analysis Thread\footnote{\url{https://heasarc.gsfc.nasa.gov/docs/nicer/analysis_threads/}}. We employed the \texttt{nicerl2} task to we generated cleaned event files. While \texttt{nicerl3-lc} was used to generate light curves, \texttt{nicerl3-spect} was used to generate source and background spectra ($0.3-10\,\mathrm{keV}$ energy range) as well as their associated response files. We chose the \textsc{scorpeon} model---with file output format---for the background spectra in all cases.

\section{Data Analysis and Results} \label{sec:three}
\begin{figure}
\includegraphics[width=.50\textwidth, angle=0, trim={3cm 1cm 2cm 0.5cm}]{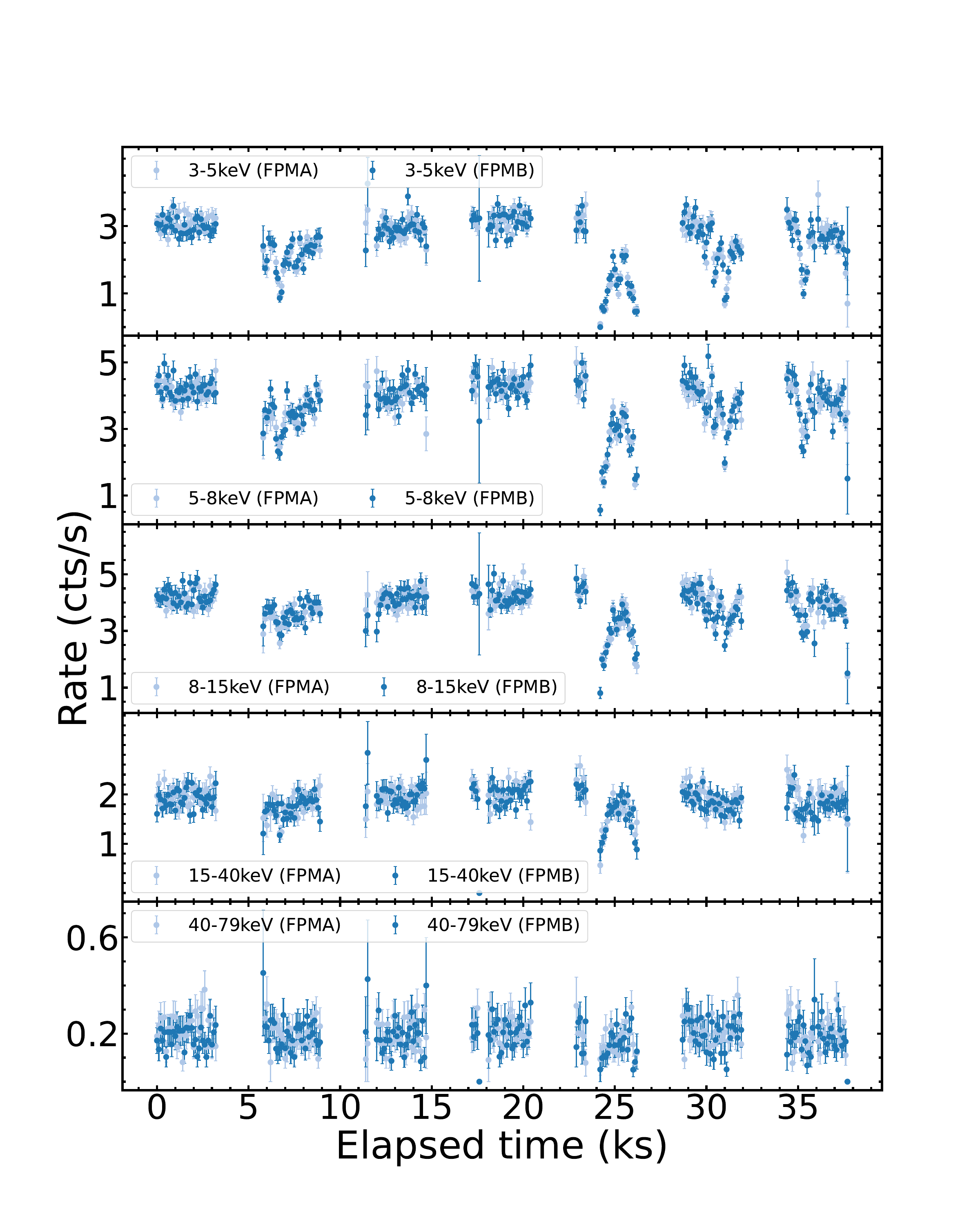}
\caption{Energy-resolved light curve for Epoch 5.}
\label{fig:epoch5_e-resolved_lc}
\end{figure}
%%%%
Figure \ref{fig:lc_nustar} shows the NuSTAR light curves from the six epochs. The source flux is fairly steady over the first three epochs, although Epoch 3 shows hints of some modulation. By Epoch 4, the overall flux has dropped by about $20\%$. The Epoch 5 observation, separated from Epoch 4 by about $83\,\mathrm{ks}$, is peculiar because it shows clear evidence of intensity dips. It is worth noting that the steady segments of the Epoch 5 light curve have about the same average flux as in Epoch 4. This is the first observed instance of unambiguous dipping behavior, attributable to obscuration, in IGR~J17091-3624. The observation shows a series of five dipping episodes in X-ray intensity.
%lasting between $\sim1-5\,\mathrm{ks}$. 
During the deepest dip, the count rate dropped by about $85\%$. As shown in Fig. \ref{fig:epoch5_e-resolved_lc}, the fractional decrease in intensity during the dips is greatest at low energies, resulting in a hardening of the X-ray spectrum, and in line with the case for an obscurer as the origin of the dips. Structured repeated variability patterns are not evident in any of the light curves. This is not surprising since past observations of the source that showed these exotic variability behaviors were mostly in the soft state or in a state with significant disk contribution.

For spectral analysis, the NuSTAR data were grouped to have a minimum of 40 counts per spectral bin, using the ``optmin'' flag in \texttt{ftgrouppha} \citep{2016A&A...587A.151K}, to ensure sufficient counts in every spectral bin. Spectral analysis was performed in \textsc{xspec} \citep{1996ASPC..101...17A} v12.13.0c using the chi-squared statistic. We modeled line-of-sight photoelectric absorption in the interstellar medium with \texttt{TBabs}, using the cosmic abundances of \citet{2000ApJ...542..914W} and the cross sections of \citet{1996ApJ...465..487V}. The hydrogen column density $N_{\rm H}$ was kept fixed at $1.1\times10^{22}\,\mathrm{cm^{-2}}$ 
%based on previously reported values
\citep[e.g.,][]{2011A&A...533L...4R, 2024ApJ...963...14W}. In all cases, the errors were computed in the confidence interval $90\%$ for one parameter of interest. The values reported in Tables \ref{tab:mo_table} and \ref{tab:relxillcp_dip+persistent} were obtained using MCMC implemented in \textsc{xspec}, employing the Goodman–Weare algorithm. For each run, we used a total chain length of $6\times10^{7}$ for $50$ walkers, with a burn-in phase of length $10^{6}$.
%of which the first $10^{6}$ steps were discarded. 
These values were chosen after a number of tests to ensure convergence. 
%\textbf{We assessed convergence by estimating the autocorrelation length per walker for varying chain length runs starting at $2\times10^{6}$. For chain length of $4\times10^{6}$, the run length is at least four times the autocorrelation length per walker for each parameter.}
Convergence was assessed using trace plots, autocorrelation analysis, and effective sample size over multiple chain length runs, starting at $2\times10^{6}$ up to $7\times10^{7}$. From chain length of $\sim4\times10^{7}$ and higher, the autocorrelation length remains fairly steady and does not increase with chain length.

\subsection{Broadband Spectra} \label{subsec:three-one}
\begin{figure*}%[ht!]
\includegraphics[width=1.0\textwidth, angle=0, trim={2cm 1cm 2cm 1cm}]{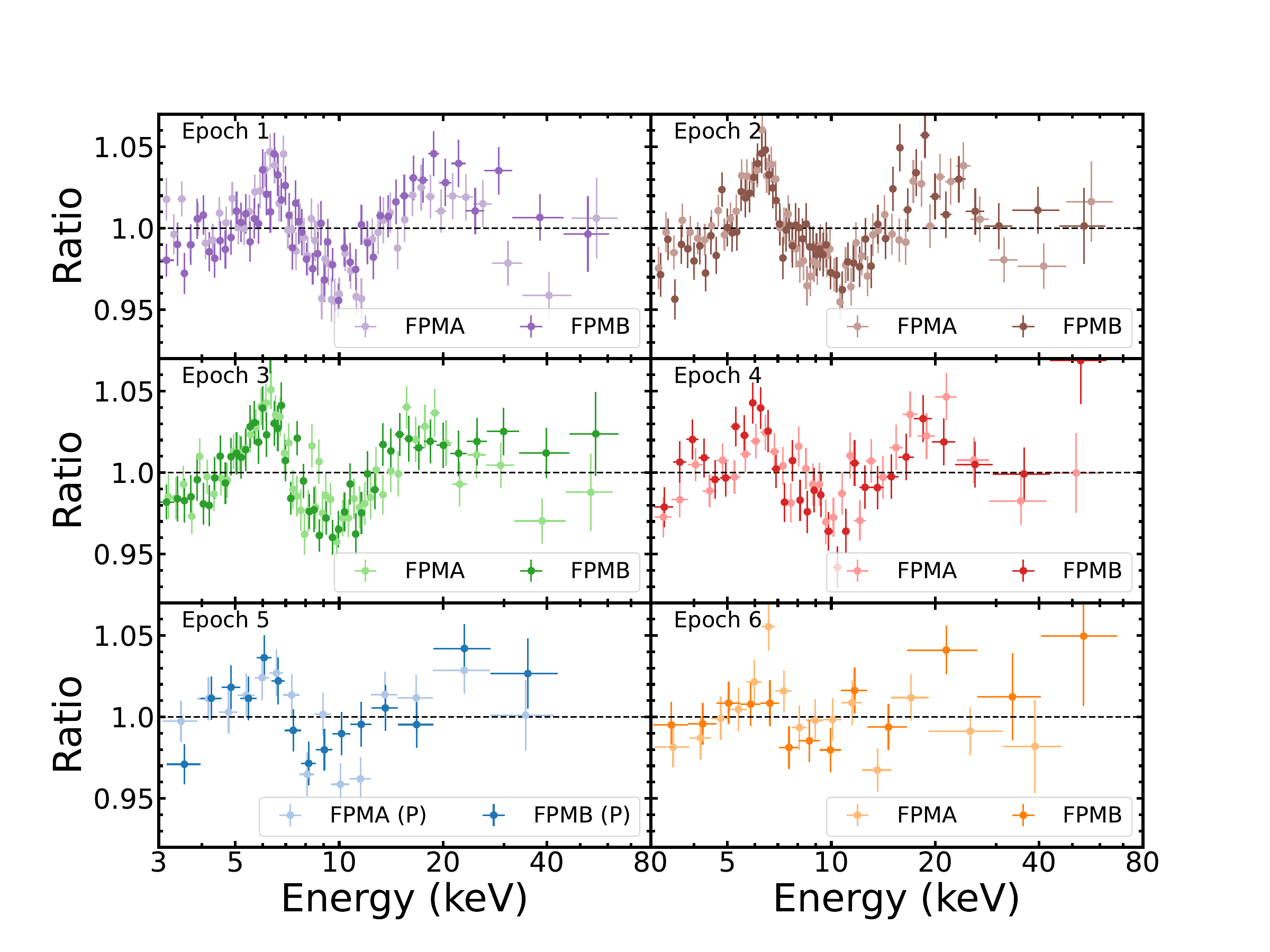}
\caption{Comptonized disk blackbody model fit to data from all six epochs using the model \texttt{cons*TBabs*simplcut*diskbb}. Relativistic reflection features are evident in all six epochs. For Epoch 5, only the persistent spectra are plotted.} 
\label{fig:epoch1-6_reflection_spec}
\end{figure*}

\begin{figure}%[ht!]
\includegraphics[width=.50\textwidth, angle=0, trim={2cm 1cm 2cm 1cm}]{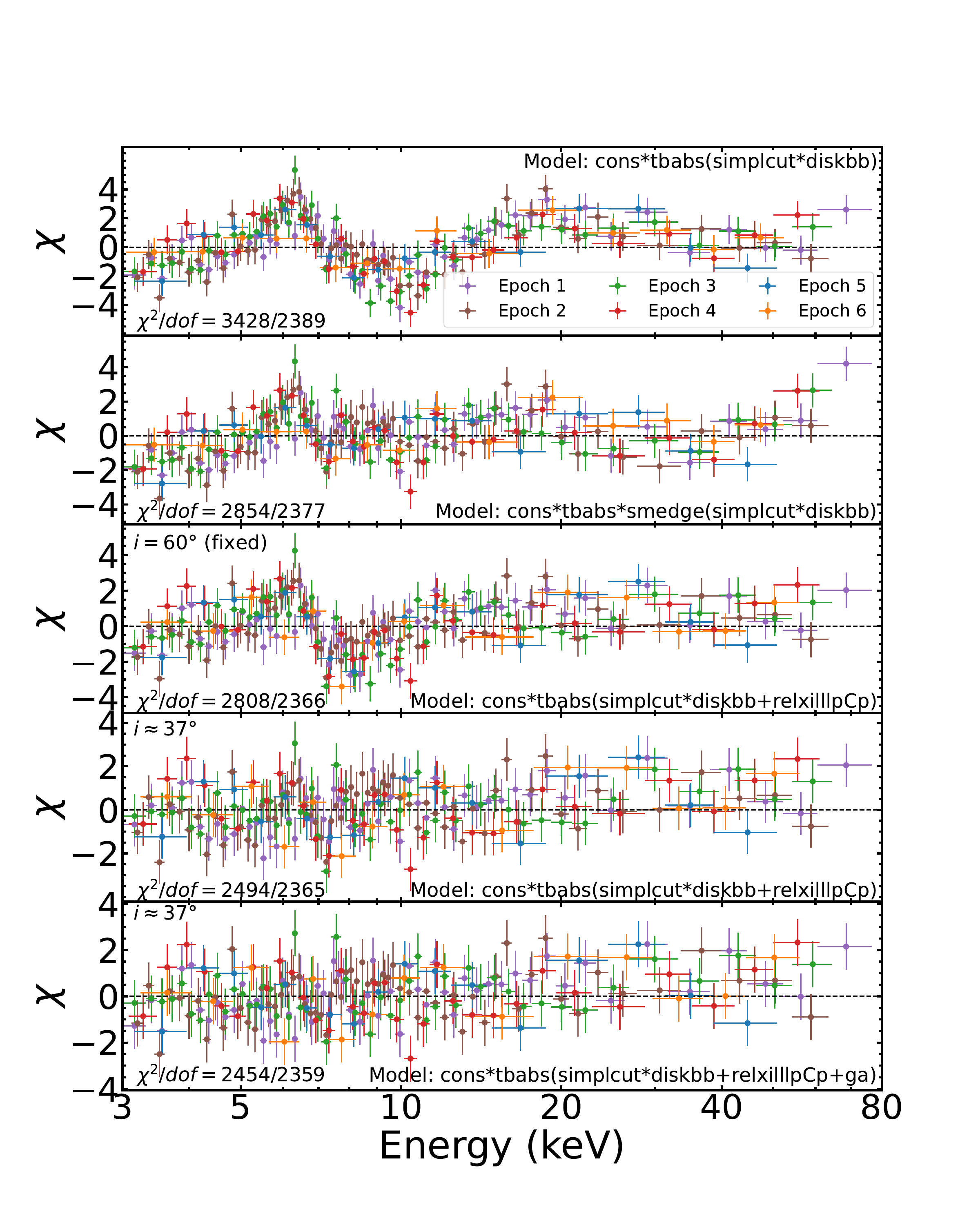}
\caption{Residuals from joint fit to the data from Epochs 1-6 using both phenomenological and physical models. The data points shown are from FPMB. In all cases, only data from the persistent spectra are considered for Epoch 5.} 
\label{fig:epoch1-5_relxilllpcp}
\end{figure}

As a first step, we fit an absorbed Comptonized-disk black body model to the NuSTAR spectra of each of the six epochs. For Epoch 5, spectra were generated for the dipping and the non-dipping or persistent intervals separately (see Fig. \ref{fig:lc_nustar}). As evident in Fig. \ref{fig:epoch1-6_reflection_spec}, residuals from fits to the individual spectra of all six epochs show significant relativistic reflection signatures---broadened neutral iron K-shell emission line near $6.4\,\mathrm{keV}$, iron K-edge absorption near $10\,\mathrm{keV}$ and a Compton hump peaking near $20\,\mathrm{keV}$. 
A careful inspection of Fig. \ref{fig:epoch1-6_reflection_spec}, as well as the first four panels of Fig. \ref{fig:epoch1-5_relxilllpcp}, 
reveals a potential narrow absorption line near $7\,\mathrm{keV}$, most apparent in Epochs 1 and 2, where the spectra benefit from higher count rates and signal-to-noise ratio. 
This feature appears weak in the individual spectra and superimposed on the broader iron-K edge structure, making it difficult to isolate. 
To fit for the broad absorption and to enhance the visibility of the narrow line, we included a \texttt{smedge} component \citep{1994PASJ...46..375E} with a fixed index ($-2.67$) and width ($7\,\mathrm{keV}$). The second panel of Fig. \ref{fig:epoch1-5_relxilllpcp} shows the joint-fit residuals after this addition, with the narrow absorption line emerging more clearly.
%%%%
%%
%%%
%%%%
To model the relativistic reflection features, the spectra from all six epochs were subsequently jointly fitted with the model \texttt{cons*TBabs(simplcut*diskbb+relxillCp)}. \texttt{simplcut} is an empirical Comptonization model that self-consistently produces a power law through the scattering of seed disk photons \citep{2009PASP..121.1279S, 2017ApJ...836..119S} whereas \texttt{diskbb} models the spectrum from an accretion disk consisting of multiple blackbody components \citep[e.g.,][]{1984PASJ...36..741M, 1986ApJ...308..635M}. \texttt{relxillCp}
%, on the other hand, 
is one of the flavors of \texttt{relxill}---a family of state-of-the-art relativistic reflection models \citep{2014MNRAS.444L.100D, 2014ApJ...782...76G}. During fitting, the inclination was initially kept frozen at $60\degree$, considering that the source may be highly inclined based on features typical of high-inclination systems, like disk winds, that have been seen in its spectra from past observations \citep[e.g.,][]{2012ApJ...746L..20K} as well as its similarity to GRS~1915+105. Because $R_{in}$, the inner radius of the disk from \texttt{relxillCp}, tends to be degenerate with the spin, its value was kept fixed at the innermost stable circular orbit (ISCO) while the spin parameter $a_*$ was allowed to vary freely. Both the photon index $\Gamma$ and the corona temperature $kT_{e}$ were tied between the components \texttt{simplcut} and \texttt{relxillCp} for each individual epoch. The reflection fraction parameter $R_{f}$ was set at -1 so that only the reflected spectrum is computed. This gave $\chi^2/dof=2803/2388$, with all spectral parameters tied except the normalizations of \texttt{diskbb} and \texttt{relxillCp} that were allowed to vary freely for the spectra from each epoch.
%, implying that the reflection spectra did not evolve significantly over the course of these five observations. 
Untying \texttt{diskbb} temperature did not improve the fit in any appreciable way whereas untying both $\Gamma$ and $kT_{e}$ improved the fit significantly with $\Delta\chi^{2}/\Delta dof=335/10$, giving $\chi^{2}/dof=2468/2378$. $\Gamma$ showed a decreasing trend from Epochs 1 to 6 as luminosity drops. The corona temperature is fairly well constrained for all six epochs, with best-fit values between $\sim25\,\mathrm{keV}$ and $\sim155\,\mathrm{keV}$. The scattered fraction $f_{scat}$ from \texttt{simplcut} quantifies the strength of the Compton power-law component relative to the disk. Its value had to be tied among all data groups, as it is not strongly constrained for the individual data groups. This is likely because data covering the lower-energy band, where the disk black-body emission contributes significantly, are missing. 
%The \texttt{diskbb} temperature is low, at $\sim0.1\,\mathrm{keV}$, not unexpected since the source remained in the hard state through the outburst. 
The inner emissivity index $q_{1}$ is not constrained, instead it is anchored at its maximum value of 10. 
%An absorption feature appears at $\sim7.2\,\mathrm{keV}$ and tends to disappear when the inclination is thawed, with best-fit value $\sim33\degree$. Thus, even though an absorption line has been reported at a similar energy for IGR~17091-3624 in the past, it is not clear if the feature is real in the present data as it is sensitive to the choice of values for inclination and/or $q_1$. 
The best-fit spin value is $0.94\pm{0.01}$, consistent with previous claims for the source \citep[e.g.,][]{2012ATel.4382....1R}, although the sensitivity of the fits to the spin value is weak. 
%Removing the Comptonized \texttt{diskbb} model, and fitting exclusively with \texttt{relxillCp} equally yielded a good fit, giving $\chi^2/dof=2112/2028$ when both $\Gamma$ and $kTe$ are untied among the data groups of the six epochs. Here, the reflection fraction is a free parameter, although tied among the spectra from the six epochs. The trend of $\Gamma$ with luminosity is consistent with that of the previous fit. The best-fit parameters are also consistent. The disk temperature is expected to remain very low in the hard state, significantly lower than the low-energy limit of NuSTAR, thus it is not unexpected that the NuSTAR spectra would return a good fit even without a disk blackbody component. The inner emissivity index remains unconstrained, anchored at its upper limit.
\begin{table*}[ht!]
\caption{Best-fit parameter values from joint model fits to the spectra of all six epochs.} % title of table
\centering % used for centering table
%\resizebox{\textwidth}
%\small
\footnotesize
\begin{tabular}{lccccccc} % %centered columns(2)
%\hline\hline %inserts double horizontal lines
%\textbf{Model/Parameter} \& \textbf{Best-fit values} \\ [0.5ex] % inserts table
%heading
\hline\hline % inserts single horizontal line
Component & parameter & Epoch 1 & Epoch 2 & Epoch 3 & Epoch 4 & Epoch 5 (P) & Epoch 6\\
\hline
Scale factor & FPMA/B & $1.007^{+0.004}_{-0.005}$ & $1.01\pm{0.01}$ & $1.00\pm{0.01}$ & $1.01\pm{0.01}$ & $1.01\pm{0.01}$ & $0.99\pm{0.01}$\\
\hline
TBabs & $N_{\rm{H}}~(10^{22}\,\mathrm{cm^{-2}})$ & \multicolumn{6}{c}{1.1} \\
%& $O~\rm{abund}.~(solar)$ & $1.32^{+0.04}_{-0.04}$ & $1.25^{+0.03}_{-0.03}$ & $1.38^{+0.03}_{-0.03}$ & $1.36\pm{0.04}$\\
%& $Fe~\rm{abund}.~(solar)$ & $1.6^{+0.1}_{-0.1}$ & $1.59^{+0.09}_{-0.09}$ & $1.13^{+0.07}_{-0.07}$ & $1.12^{+0.08}_{-0.08}$\\
%& $redshift$ & $0$(f) & $0$(f) & $0$(f) & $0$(f) & $0$(f) & $0$(f)\\
%Gal. abs. & $N_{\rm{H}} (10^{21}\,cm^{-2})$ & 0.00(f) & $16.48\pm{1.18}$ & $-$ & $-$ & $-$ & $-$\\
\hline
simplcut & $\Gamma$ & $1.65^{+0.02}_{0.04}$ & $1.63^{+0.02}_{-0.03}$ & $1.60^{+0.03}_{-0.05}$ & $1.59^{+0.02}_{-0.05}$ & $1.58\pm{0.03}$ & $1.56^{+0.03}_{0.05}$\\
& $f_{\rm{scat}}$ & \multicolumn{6}{c}{$0.87$} \\
& $Refl_{\rm{frac}}$ & \multicolumn{6}{c}{1} \\
& $kTe\, (\rm{keV})$ & $25^{+5}_{-2}$ & $23^{+6}_{-3}$ & $23^{+6}_{-2}$ & $30^{+6}_{-4}$ & $29^{+8}_{-5}$ & $155^{+69}_{-68}$\\
\hline
diskbb & $T_{in}$ (keV) & $0.121^{+0.001}_{-0.008}$ & $0.12\pm{0.01}$ & $0.12\pm{0.01}$ & $0.12\pm{0.01}$ & $0.12^{+0.02}_{-0.01}$ & $0.12^{+0.02}_{-0.01}$\\
& norm ($\times10^{4}$) & $6.7\pm{0.2}$ & $5.5^{+2.0}_{-1.2}$ & $4.9^{+1.0}_{-1.2}$ & $4.4^{+0.8}_{-1.4}$ & $3.9^{+2.1}_{-1.4}$ & $1.4^{+0.4}_{-0.6}$\\
\hline
relxilllpCp & $i~({\degree})$ & \multicolumn{6}{c}{$37^{+4}_{-5}$\degree} \\
& $a_*$ & \multicolumn{6}{c}{$0.947^{+0.045}_{-0.224}$} \\
& $R_{in}~(\rm{ISCO})$ & \multicolumn{6}{c}{$1$} \\
& $R_{out}~(\rm{r_{g}})$ & \multicolumn{6}{c}{$400$} \\
& $h~(\rm{r_{g}})$ & $4.4^{+0.9}_{-0.6}$ & $3.2^{+0.7}_{-0.3}$ & $3.0^{+0.7}_{-0.4}$ & $3.3^{+0.7}_{-0.4}$ & $4.2^{+0.6}_{-0.3}$ & $13.9^{+3.7}_{-3.0}$ \\
%& $q_{1}$ & \multicolumn{5}{c}{$0.052^{+0.003}_{-0.003}$} & $-$\\
%& $q_{2}$ & \multicolumn{5}{c}{$0.052^{+0.003}_{-0.003}$} & $-$\\
%& $redshift$ & $0$(f) & $0$(f) & $0$(f) & $0$(f) & $0$(f) & $0$(f)\\
%& $\Gamma^{\dagger}$ & $1.633$ & $1.631$ & $1.584$ & $1.573$ & $1.570$ & $1.544$\\
& log~[$\xi/\mathrm{erg\,cm\,s^{-1}}$] & $2.9^{+0.6}_{-0.2}$ & $3.0^{+0.4}_{-0.2}$ & $3.1^{+0.4}_{-0.3}$ & $3.2^{+0.4}_{-0.3}$ & $3.2\pm{0.3}$ & $2.8^{+0.7}_{-1.7}$\\
& log~[$N/\mathrm{cm^{-3}}]$ & \multicolumn{6}{c}{$19^{+1}_{-2}$} \\
& $A_{Fe}$ (solar) & \multicolumn{6}{c}{$1.1^{+0.3}_{0.2}$} \\
%& $kTe^{\dagger}\, (\rm{keV})$ & $23$ & $24$ & $22$ & $30$ & $30$ & $100$\\
& $Refl_{\rm{frac}}$ & \multicolumn{6}{c}{$-1$} \\
& $\mathrm{norm}_{\rm{relxilllpcp}}~(10^{-6})$ & $1959^{+27}_{-43}$ & $3493^{+1423}_{-1134}$ & $4398^{+1754}_{-754}$ & $2059^{+330}_{-170}$ & $1478^{+344}_{-465}$ & $130^{+48}_{-22}$\\
\hline
gauss & ${E_{\rm{abs}}}$ (keV) & \multicolumn{6}{c}{$7.13^{+0.14}_{-1.73}$} \\
& $\sigma$ (keV) & \multicolumn{6}{c}{$0.01$} \\
& norm $(10^{-5})$ & \multicolumn{6}{c}{$-1.7^{+0.2}_{-0.4}$} \\
\hline
unabsorbed flux & ${2-10\,\mathrm{keV}}~(10^{-10}\,\mathrm{erg\,cm^{-2}\,s^{-1}})$ & $4.55\pm{0.01}$ & $4.35\pm{0.01}$ & $4.17\pm{0.01}$ & $3.40\pm{0.01}$ & $3.30\pm{0.2}$ & $1.02\pm{0.01}$\\
\hline
relxilllpCp flux & $0.1-100\,\mathrm{keV}~(10^{-10}\,\mathrm{erg\,cm^{-2}\,s^{-1}})$ & $4.5\pm{0.2}$ & $4.4\pm{0.2}$ & $4.6\pm{0.3}$ & $2.8\pm{0.4}$ & $3.2^{+0.6}_{-0.5}$ & $0.8\pm{0.1}$\\
unabsorbed flux & ${0.1-100\,\mathrm{keV}}~(10^{-10}\,\mathrm{erg\,cm^{-2}\,s^{-1}})$ & $22.6\pm{0.1}$ & $21.5\pm{0.1}$ & $20.6\pm{0.1}$ & $17.6\pm{0.1}$ & $17.3\pm{0.1}$ & $5.96\pm{0.04}$\\
Reflection strength & $R_{str}~(\%)$ & $20\pm{1}$ & $20\pm{1}$ & $22\pm{2}$ & $16\pm{2}$ & $18\pm{3}$ & $13\pm{2}$\\
\hline    
%gabs & ${E_{\rm{abs}}}$ (keV) & $0.6^{+0.1}_{-0.2}$ & $1.2\pm{0.1}$ & $-$ & $-$\\
%& $\sigma$ (keV) & $2.3^{+0.2}_{-0.1}$ & $2.0\pm{0.1}$ & $-$ & $-$\\
%& strength $(10^{-2})$ & $87^{+13}_{-17}$ & $264\pm{1}$ & $-$ & $-$\\
%edge & $E_{\rm{edge}}$ (keV) & $-$ & $-$ & $-$ & $0.39\pm{0.01}$\\
%& $\tau_{\rm{max}}$ & $-$ & $-$ & $-$ & $0.6\pm{0.1}$\\
$\chi^2/dof$ & & \multicolumn{6}{c}{2454/2359} \\ 
%Flux ($\mathrm{erg\,cm^{-2}\,s^{-1}}$) & $6.66\times10^{-12}$\\
%$L$ ($\mathrm{erg\,s^{-1}}$) & $1.14\times10^{43}$\\
\hline
\hline %inserts single line
\end{tabular}
\label{tab:mo_table} % is used to refer this table in the text

{Note: Parameters without uncertainties were kept frozen at the quoted values. $f_{scat}$ is poorly constrained, it is thus frozen at its fiducial best-fit value. The fluxes were estimated, not from the chains, but using \texttt{cflux} with the best-fit model from \textsc{xspec}.}
\end{table*}
%%%%

The emissivity profile is a vital component of reflection models because it quantifies the radial dependence of the intensity of the reflected emission. When the primary source is close to the black hole, light-bending effects will concentrate its radiation on the inner parts of the disk, which can result in a high value of $q_{1}$. If the corona geometry is known, the emissivity profile can be self-consistently computed. This is implemented in the \texttt{relxilllpCp} flavor of the \texttt{relxill} family of models. \texttt{relxilllpCp} assumes the primary source to be point-like with a lamp-post geometry, and on the rotational axis of the black hole. Thus, the emissivity depends on the height of the source above the black hole and, potentially, also on its velocity $\beta$ along this axis \citep{2013MNRAS.430.1694D}. Although the lamppost is an idealized geometry, it allows for the determination of several key parameters such as the proximity of the primary source to the black hole and the relative strength of the direct and reflected components.
We therefore replaced \texttt{relxillCp} with \texttt{relxilllpCp} from the model described above, that is, \texttt{cons*TBabs(simplcut*diskbb+relxilllpCp)}. We kept the inclination frozen at $60\degree$, $R_{in}$ at the ISCO, the corona height $h$ tied among all data groups and $\beta$ set to zero. This gave $\chi^2/dof=2808/2366$. The residual plot shows features consistent with an absorption line at $\sim7.2\,\mathrm{keV}$ and a broad positive feature near $6\,\mathrm{keV}$ (see Fig. \ref{fig:epoch1-5_relxilllpcp}, middle panel). Unfreezing the inclination significantly improved the fit, with $\Delta\chi^{2}=314$ for one additional free parameter, giving $\chi^2/dof=2494/2365$. 
The best-fit inclination value was $\sim37$\degree. With this, as the fourth panel of Fig. \ref{fig:epoch1-5_relxilllpcp} shows, most of the residuals between $5-10\,\mathrm{keV}$ were accounted for. The disk black body temperature $T_{in}$ is comparable among all six spectral groups, suggesting that the disk reflection properties may not have evolved appreciably over the course of the observations. Most of the spectral evolution appears to have occurred in the corona alone, especially for the first five epochs. Although, the fact that NuSTAR's energy coverage does not go below $3\,\mathrm{keV}$ means that any constraints on disk temperature is likely weak at best. 
%although an absorption feature still appears visible around $7.2\,\mathrm{keV}$. 
Adding a Gaussian line---width kept at $0.01\,\mathrm{keV}$---to fit for any absorption line residuals slightly improved the fit further, with $\Delta\chi^{2}/\Delta dof=29/2$, giving $\chi^2/dof=2465/2363$. Untying $h$ further improved the fit marginally, giving $\chi^2/dof=2454/2359$ ($f_{scat}$ has been frozen since it is poorly constrained). The best-fit inclination value was $37^{+4}_{-5}$\degree. The best-fit line energy was $7.13^{+0.14}_{-1.73}\,\mathrm{keV}$. 
The line appears to be detected above the $3\sigma$ confidence level (estimated by dividing the \texttt{Gauss} normalization by its negative error). 
The fit parameters are shown in Table \ref{tab:mo_table}. Strong absorption features from highly ionized iron are signatures of powerful disk winds and are not commonly detected from BHXBs in the low/hard state. Associating the line with the H-like Fe \textsc{xxvi} absorption line would imply an upper limit of $6000\,\mathrm{km\,s^{-1}}$ on the velocity shift, at the $90\%$ confidence interval.
%, it will correspond to a velocity shift of $8000^{+7000}_{-23000}\,\mathrm{km\,s^{-1}}$. 

As Fig. \ref{fig:lc_nustar} shows, during Epoch 6, the count rate has dropped significantly. It is worth mentioning that the Comptonized disk blackbody model, \texttt{cons*TBabs(simplcut*diskbb)}, alone tends to equally reproduce the broadband spectra appreciably well for Epoch 6, giving $\chi^2/dof=373/355$ and $\Gamma=1.562^{+0.002}_{-0.030}$. An absorbed cutoff power law also provided a good fit to the spectra, with $\chi^2/dof=382/357$ and $\Gamma=1.47\pm{0.01}$.
%Relativistic reflection signatures, prominent in the first five epochs, are largely missing, as the residual plot of Fig. \ref{fig:epoch1-6_reflection_spec} shows. Also, the spectra is significantly harder, with $\Gamma=1.56\pm{0.01}$. 
This may indicate that the disk has receded further, becoming more truncated than in previous epochs. The corona height $h$ is significantly higher, at $13.9^{+3.7}_{-3.0}\,\mathrm r_{\rm g}$, and the estimated reflection strength $R_{str}$ is significantly lower for Epoch 6, on average, compared to other epochs (see Table \ref{tab:mo_table}), a further confirmation that the spectrum has evolved considerably during this observation. This is characteristic of the low/hard state of BHXBs, prior to the return to quiescence.
%\begin{figure*}
%\includegraphics[width=.30\textwidth, angle=0, trim={3cm 1cm 2cm 1cm}]{nu-epoch5_energy-resolved_lc_igr.pdf}
%\includegraphics[width=.30\textwidth, angle=0, trim={2cm 1cm 3cm 1cm}]{simplcut-diskbb+relxillcp_spec_dip+persistent_freeze.pdf}
%\includegraphics[width=.30\textwidth, angle=0, trim={2cm 1cm 3cm 1cm}]{xstar_simplcut-diskbb+relxillcp_spec_dip+persistent.pdf}
%\caption{\textit{Left}: Energy-resolved light curve for Epoch 5. \textit{Middle}: Best-fit to the persistent spectra of Epoch 5 using the model \texttt{cons*TBabs(simplcut*diskbb+relxillCp)} with the dip spectra overlaid. \textit{Right}: Overall best-fit to the joint persistent and dip spectra employing the model \texttt{cons*TBabs*XSTAR(simplcut*diskbb+relxillCp)}.}
%\label{fig:epoch5_nustar}
%$\end{figure*}
%%%%
\subsection{Dip vs. Persistent Spectra} \label{subsec:three-two}
\begin{figure*}
\includegraphics[width=.50\textwidth, angle=0, trim={2cm 1cm 1.8cm 1cm}]{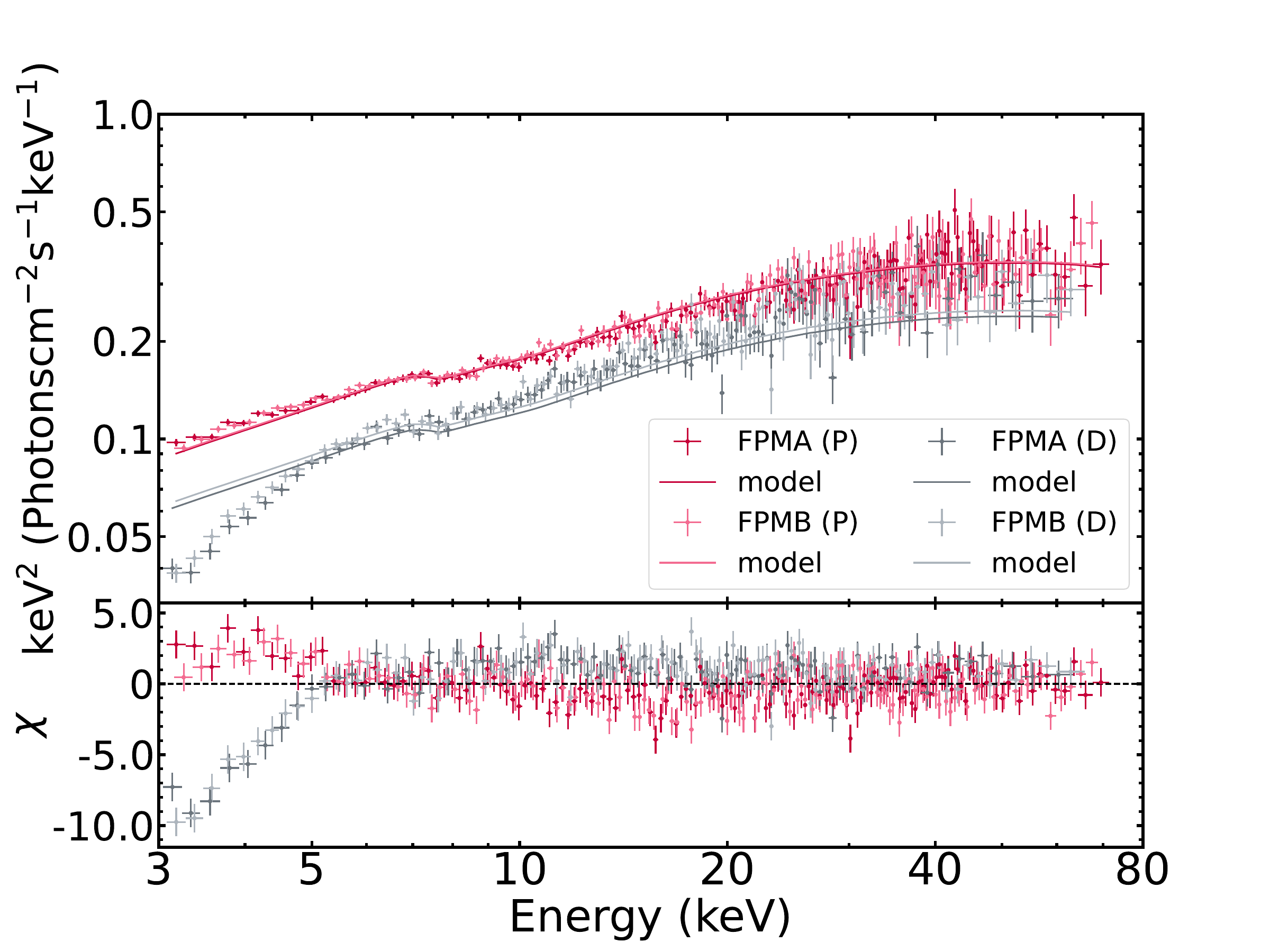}
\includegraphics[width=.50\textwidth, angle=0, trim={1.80cm 1cm 2cm 1cm}]{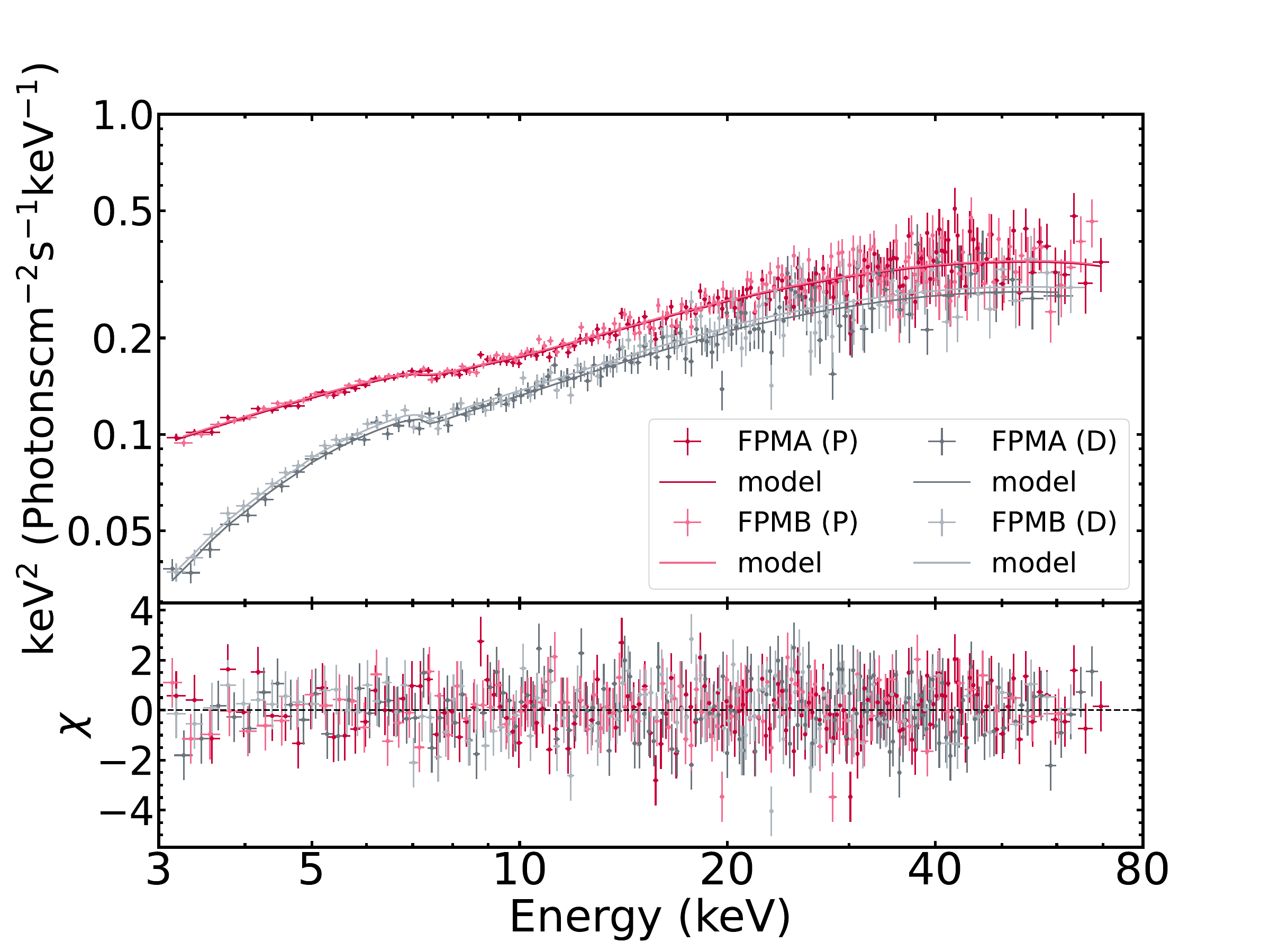}
\caption{\textit{Left}: Best-fit to the persistent spectra of Epoch 5 using the model \texttt{cons*TBabs(simplcut*diskbb+relxillCp)} with the dip spectra overlaid. \textit{Right}: Overall best-fit to the joint persistent and dip spectra employing the model \texttt{cons*TBabs*XSTAR(simplcut*diskbb+relxillCp)}.}
\label{fig:epoch5_nustar}
\end{figure*}
%%%%
As shown in Fig. \ref{fig:lc_nustar}, during Epoch 5, the light curve showed intervals of recurrent flux dips indicative of possible obscuration along the line-of-sight to the X-ray source. While this has been observed in a handful of high-inclination systems, this is the first time it is seen in IGR~J17091-3624. The count rates in the light curve of Epoch 4 and the persistent interval in Epoch 5 are fairly consistent with each other, and their spectral shapes are equally similar. To probe the properties of the obscurer, we generated spectra for intervals exclusively covering the dips (dip spectra) and for intervals excluding the dips (persistent spectra), and then we carried out a joint spectral fit. Fitting the persistent spectra with the model \texttt{cons*TBabs(simplcut*diskbb+relxillCp)} yielded an acceptable fit with $\chi^2/dof=333/312$. However, adding the dip spectra and tying its parameters to the best-fit parameters from the persistent spectra yielded a poor fit,
giving $\chi^2/dof=1587/546$. 
The residual plot of Fig. \ref{fig:epoch5_nustar} (left) shows that in addition to the drop in flux (about $20\%$ at all energies) during the dips, the broadband spectrum is also significantly modified especially below $\sim5\,\mathrm{keV}$. The inclusion of an extra \texttt{TBabs} model, exclusively for the dip spectra  (i.e. its $N_{\rm H}$ is set to zero for the persistent spectra) returned an excellent fit, with $\Delta\chi^{2}=1024$, giving $\chi^2/dof=563/545$. The best-fit $N_{\rm H}$ for the obscurer responsible for the dips is $\sim10^{23}\,\mathrm{cm^{-2}}$. 

We subsequently replaced the added \texttt{TBabs} with an \texttt{XSTAR}-generated table model \citep{2001ApJS..133..221K} to fit for the absorption imprinted on the dip spectra. In generating the \texttt{XSTAR} grid, we used an input spectral file obtained from fitting a simple Comptonized disk blackbody to the dip spectra from NuSTAR. We assumed a gas density of $10^{14}\,\mathrm{cm^{-3}}$ and a source luminosity of $10^{38}\,\mathrm{erg\,s^{-1}}$ \citep[e.g.,][]{2024ApJ...977...26A}. The grid covered the parameter space of $10^{18}\,\mathrm{cm^{-2}}\leq N_{\rm H}\leq10^{24}\,\mathrm{cm^{-2}}$ and $-1\leq \mathrm{log}~[\xi^{xstar}/\mathrm{erg\,cm\,s^{-1}}] \leq 4$. The complete model is now \texttt{cons*TBabs*XSTAR(simplcut*diskbb+relxillCp)}. During fitting, we allowed $N_{\rm H}^{xstar}$, the column density from \texttt{XSTAR}, and the ionization parameter ${\rm{log}}[\,\xi^{xstar}/\mathrm{erg\,cm\,s^{-1}}]$ to be free for the dip spectra but kept frozen at their minimum values for the persistent spectra. $f_{scat}$ is kept frozen at its fiducial value and the corona temperature is fixed to the best fit value for epoch 5 in Table \ref{tab:mo_table}. The model returned an equally good fit, with $\chi^2/dof=567/546$. The best-fit column density and ionization parameter for the absorber are $\sim2\times10^{23}\,\mathrm{cm^{-2}}$ and log~[$\xi/\mathrm{erg\,cm\,s^{-1}}]\sim2.2$, respectively. The best-fit parameter values are listed in Table \ref{tab:relxillcp_dip+persistent}.

%%%%
\begin{table}%[H]
\caption{Best-fit parameter values from joint fit to the persistent and dip spectra of Epoch 5.} % title of table
\centering % used for centering table
\resizebox{\columnwidth}{!}{%
%\small
\begin{tabular}{llll} % %centered columns(2)
%\hline\hline %inserts double horizontal lines
%\textbf{Model/Parameter} \& \textbf{Best-fit values} \\ [0.5ex] % inserts table
%heading
\hline\hline % inserts single horizontal line
Component & Parameter & Epoch 5(p) & Epoch 5(d)\\
\hline
Scale factor & FPMA & $1$ & $0.81\pm{0.02}$\\
& FPMB & $1.01\pm{0.01}$ & $0.84\pm{0.02}$\\
%Scale factor (NICER) & & $1.037^{+0.004}_{-0.003}$\\
\hline
Gal. abs. & $N_{\rm{H}}~(10^{22}\,\rm{cm^{-2}})$ & \multicolumn{2}{c}{$1.1$}\\
\hline
XSTAR & $N_{\rm{H}}^{xstar}~(10^{22}\,\mathrm{cm^{-2}})$ & $0.0001$ & $20\pm{3}$\\
& log~[$\xi^{xstar}/\mathrm{erg\,cm\,s^{-1}}$] & $-1$ & $2.18^{+0.04}_{-0.14}$\\
\hline
simplcut & $\Gamma$ & \multicolumn{2}{c}{$1.60^{+0.03}_{-0.06}$}\\
& $f_{\rm{scat}}$ & \multicolumn{2}{c}{$0.76$}\\
& $Refl_{frac}$ & \multicolumn{2}{c}{$1$}\\
& $kTe\, (\rm{keV})$ & \multicolumn{2}{c}{$29$}\\
\hline
diskbb & $T_{in}$ (keV) & \multicolumn{2}{c}{$0.15^{+0.03}_{-0.07}$}\\
& norm $(10^{4})$ & \multicolumn{2}{c}{$2^{+19}_{-1}$}\\
\hline
relxillCp & $i\, (\rm{\degree})$ & \multicolumn{2}{c}{$44^{+6}_{-37}$}\\
& $a_*$ & \multicolumn{2}{c}{$0.95^{+0.01}_{-1.78}$}\\
& $R_{in}~(\rm{ISCO})$ & \multicolumn{2}{c}{$1$}\\
& $R_{out}~(\rm{r_{g}})$ & \multicolumn{2}{c}{$400$}\\
& $R_{br}~(\rm{r_{g}})$ & \multicolumn{2}{c}{$15$}\\
& $q_{1}$ & \multicolumn{2}{c}{$4.6^{+4.8}_{-2.9}$}\\
& $q_{2}$ & \multicolumn{2}{c}{$3$}\\
%& $redshift$ & $0$(f)\\
%& $\Gamma$ & \multicolumn{2}{c}{$1.61^{\dagger}$}\\
& log~[$\xi/\mathrm{erg\,cm\,s^{-1}}$] & \multicolumn{2}{c}{$2.83^{+1.15}_{-2.45}$}\\
& log~[$N/\mathrm{cm^{-3}}]$ & \multicolumn{2}{c}{$19^{+1}_{-4}$}\\
& $A_{Fe}$ (solar) & \multicolumn{2}{c}{$4^{+4}_{-3}$}\\
%& $kT_{e}\, (\rm{keV})$ & \multicolumn{2}{c}{$36^{\dagger}$}\\
& $Refl_{frac}$ & \multicolumn{2}{c}{$-1$}\\
& $\mathrm{norm}_{\rm{relxillCp}}~(10^{-6})$ & \multicolumn{2}{c}{$531^{+79}_{-274}$}\\
\hline
%xillverCp & $\Gamma$ & $-$\\
%& $A_{Fe}$ & $-$\\
%& $kTe (keV)$ & $-$\\
%& $log\xi (\mathrm{erg\,cm\,s^{-1}})$ & $-$\\
%& $logN (cm^{-3})$ & $-$\\
%& $redshift$ & $-$\\
%& $i (\degree)$ & $-$\\
%& $Refl_{frac}$ & $-$\\
%& $n_{rel}(10^{-4})$ & $-$\\
%gauss & ${E_{\rm{abs}}}$ (keV) & $6.74^{+0.05}_{-0.06}$\\
%& $\sigma$ (keV) & $0.05$(f)\\
%& norm $(10^{-4})$ & $-1.8^{+0.3}_{-0.5}$\\
%edge & $E_{\rm{edge}}$ (keV) & $0.41^{+0.05}_{-0.03}$\\
%& $\tau_{\rm{max}}$ & $0.21^{+0.07}_{-0.12}$\\ 
%Flux ($\mathrm{erg\,cm^{-2}\,s^{-1}}$) & $6.66\times10^{-12}$\\
%$L$ ($\mathrm{erg\,s^{-1}}$) & $1.14\times10^{43}$\\
$\chi^2/dof$ & & \multicolumn{2}{c}{567/546}\\
\hline
\hline %inserts single line
\end{tabular}
}
\label{tab:relxillcp_dip+persistent} % is used to refer this table in the text

{Note: Parameters without uncertainties were kept frozen at the quoted values. $f_{scat}$ is poorly constrained, it is thus frozen at its fiducial best-fit value.}
\end{table}
%%%%

\subsection{Timing analysis}
\begin{figure*}
\includegraphics[width=.50\textwidth, angle=0, trim={0.5cm 1cm 0cm 1cm}]{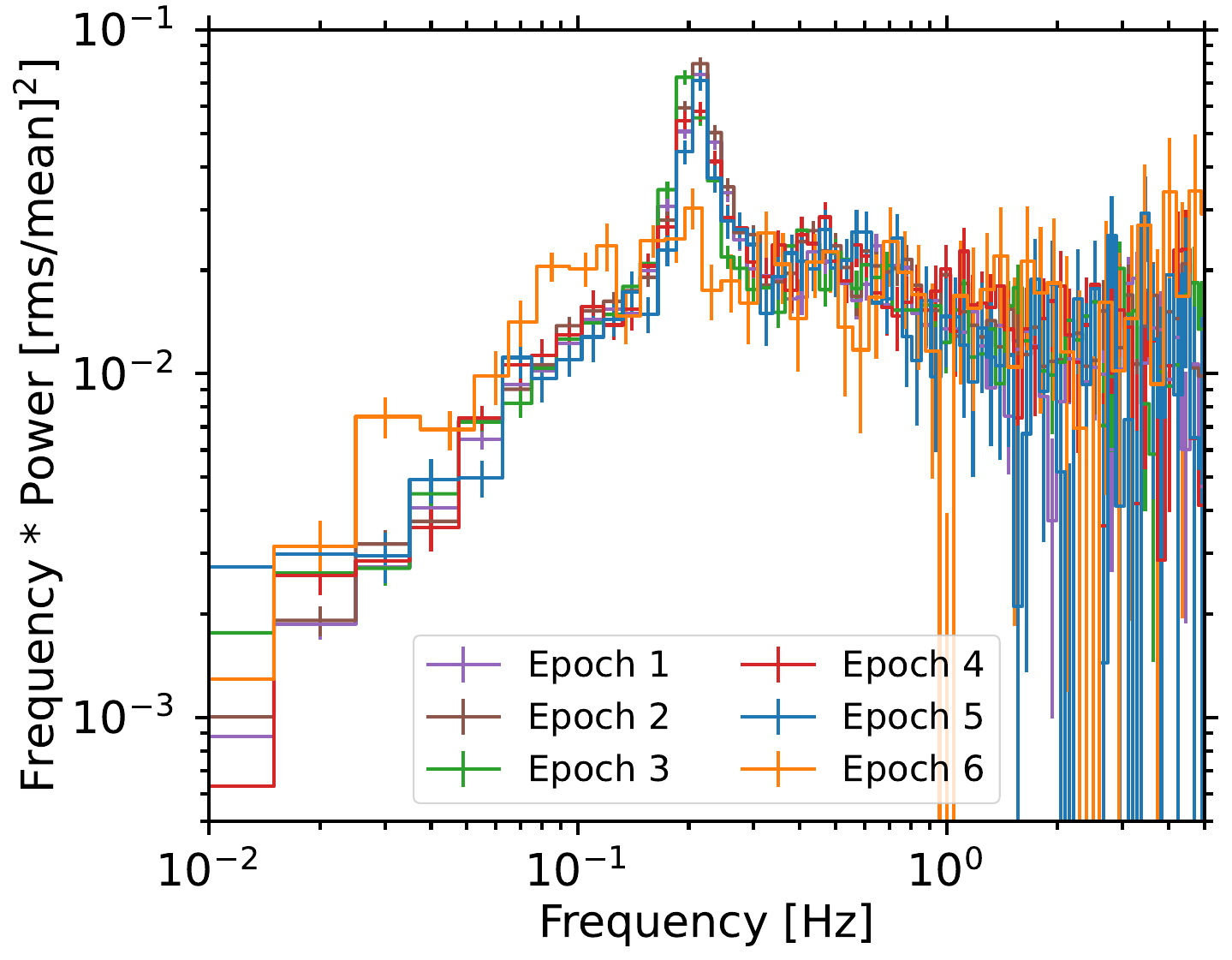}
\includegraphics[width=.50\textwidth, angle=0, trim={0cm 1cm 0.5cm 1cm}]{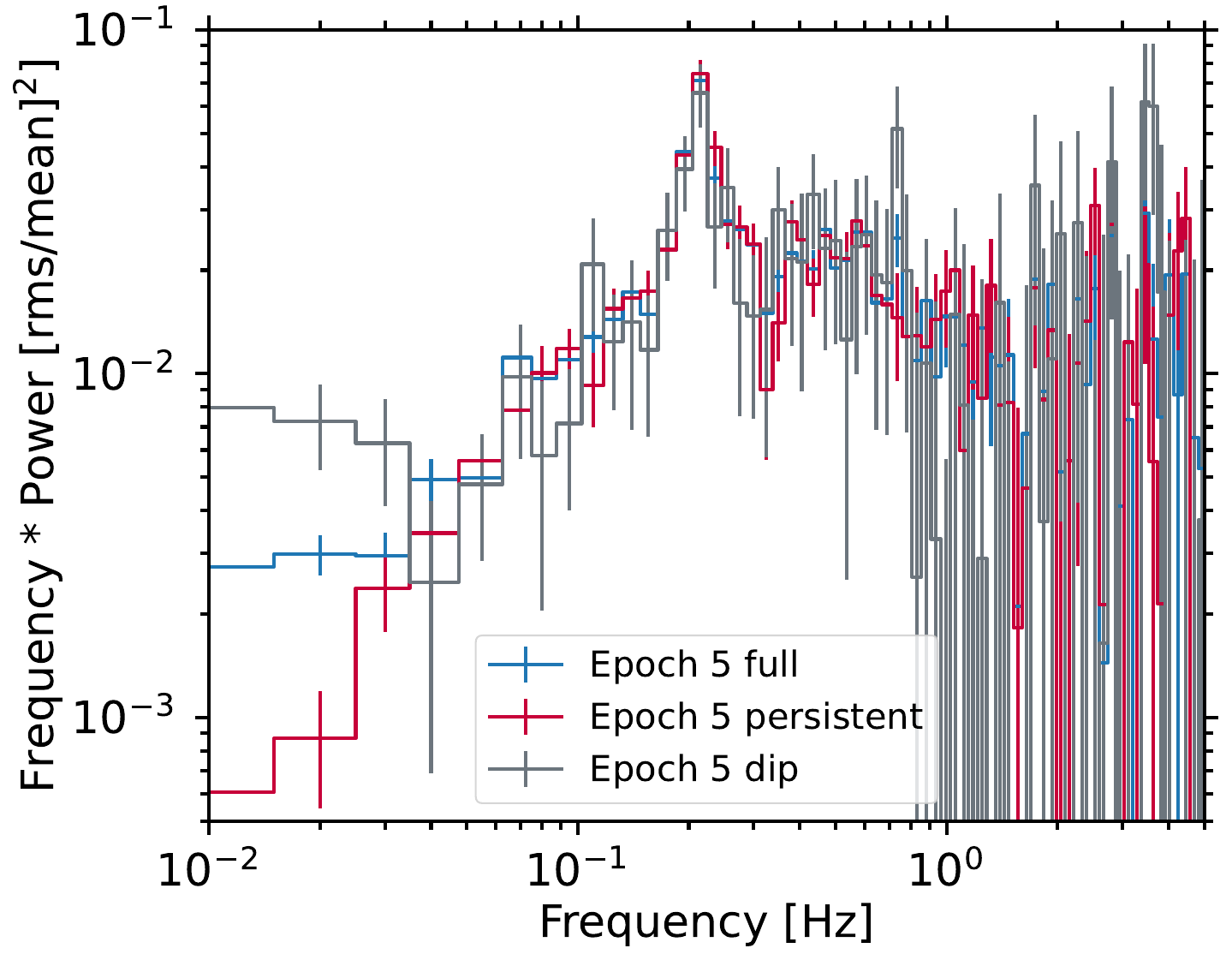}
\caption{
\textit{Left}: Power density spectra (PDS) of the six NuSTAR epochs, shown in units of fractional rms squared. Variability above $\sim2\,\mathrm{Hz}$ is consistent with zero. 
\textit{Right}: PDS of Epoch 5, extracted during the full interval, persistent phase, and dip intervals of the light curve. In both panels, the Poisson noise level was estimated by averaging the power above $100\,\mathrm{Hz}$ and subtracted across all frequencies.
}
\label{fig:PDS}
\end{figure*}
%We performed the timing analysis of all the considered epochs. All epochs exhibit fast variability detectable through Fourier analysis via the power density spectrum (PDS). \
%We utilized light curves in the full \textit{NuSTAR} energy range with a time resolution of $0.001$~s. The averaged PDS were computed using $100$~s segments and fractional root-mean-squared (rms) units, employing the \texttt{stingray} software package \citep{Huppenkothen2019, Bachetti2018}. To account for dead time effects, we applied the Fourier amplitude difference method \citep{Bachetti2018}. The high temporal resolution enabled accurate characterization of the Poisson noise component, which was modeled as a constant at frequencies above $100$~Hz and subsequently subtracted from the PDS. The left panel of Fig.~\ref{fig:PDS} presents the PDS for each epoch in this observational campaign. It is immediately evident that the first five epochs display remarkably similar PDS, each featuring a prominent quasi-periodic oscillation (QPO) centered around $\sim$0.2 Hz\footnote{We found remarkably similar frequency centroid of the QPOs, fitting each PDS epoch with Lorentian functions: 1) $ 0.211\pm 0.002$ 2) $ 0.211\pm0.002$, $ 0.200\pm 0.002$, $ 0.207^{+0.005}_{-0.003} $ and $ 0.211\pm 0.002 $}.
We carried out timing analysis for all epochs in our dataset. Each observation exhibits rapid variability, detectable through Fourier techniques using power density spectra (PDS). The analysis employed light curves extracted across the full \textit{NuSTAR} energy range with a time resolution of $0.001\,\mathrm{s}$. Averaged PDS were computed over $100\,\mathrm{s}$ segments and expressed in fractional root-mean-squared (rms) units using the \texttt{Stingray} software package \citep{Huppenkothen2019, Bachetti2018}. To correct for instrumental dead time, we applied the Fourier Amplitude Difference (FAD) method \citep{Bachetti2018}. The high temporal resolution allowed for precise modeling of the Poisson noise component, which was estimated by averaging the power above $100\,\mathrm{Hz}$ and subtracted from all frequencies.

The left panel of Figure~\ref{fig:PDS} displays the resulting PDS for all six epochs. The first five epochs show remarkably similar profiles, each dominated by a strong quasi-periodic oscillation (QPO) feature centered at approximately $0.2\,\mathrm{Hz}$. 
Table \ref{tab:psd} lists the centroid frequencies of the Lorentzian components used to model the QPOs across the different epochs. The full PDS, spanning the 0.1–500 Hz range, were first fitted with a model comprising four Lorentzian components (reduced to three in the final epoch) together with a constant term to represent the Poisson noise. The noise level was then estimated and subtracted from the PDS.
%Table \ref{tab:psd} shows the centroid frequencies
%\textbf{parameters only of the Lorentzians components responsible to fit the QPOs in the different epochs. 
%The fit of the entire PDS ranging 0.1-500 Hz was performed before subtracting the Poisson noise with a model considering 4 Lorentzian components (only 3 in the last epoch) and a constant for the noise}.
%\footnote{Centroid frequencies were derived from Lorentzian fits to each epoch's PDS.}.
%%%
\begin{table}%[H]
\caption{QPO centroid frequencies and rms amplitudes. Errors were estimated at 90\% confidence level.} % title of table
\centering % used for centering table
%\resizebox{\columnwidth}{!}{%
\begin{tabular}{c c c} % %centered columns(2)
\hline\hline %inserts double horizontal lines
\textbf{Epoch} & $\bm{{\rm QPO}_{ \rm freq} }$ \textbf{(Hz)} & $\bm{{ \rm rms} }$ \textbf{(\%)}\\ [0.5ex] %  & $\bm{R_{\rm src}}$ 
% & \textbf{(Hz)} & \textbf{(\%)}\\ [0.5ex]% & $\bm{(\arcsec)} $
%heading
\hline\hline % inserts single horizontal line
1 & $0.211\pm{0.002}$ & $27.0\pm{0.2}$ \\% & 100
%& \textit{NICER}& 3133010104 & 2021-01-23 17:26:25 & 1.2\\
%& \textit{NICER}& 3133010105 & 2021-01-23 23:39:07 & 8.6\\
2 & $0.211\pm{0.002}$ & $28.2\pm{0.2}$ \\% & 150
%& \textit{NICER}& 3133010116 & 2021-02-05 04:51:20 & 5.3\\
3 & $0.200\pm{0.002}$ & $26.9\pm{0.2}$ \\% & 100
%& \textit{NICER}& 3558010501 & 2021-02-20 03:52:17 & 1.5\\
4 & $0.207^{+0.005}_{-0.003}$ & $27.6\pm{0.3}$ \\% & 100
%& \textit{NICER}& 3558010902 & 2021-03-07 04:11:11 & 4.5\\
5 & $0.211\pm{0.002}$ & $24.5\pm{0.3}$ \\% & 100
%& \textit{NICER}& 4133010103 & 2021-03-26 03:22:39 & 6.1\\
6 & $-$ & $26.0\pm{0.6}$ \\% & 100
%& \textit{NICER}& 4133010109 & 2021-04-01 00:21:20 & 16.1\\
\hline 
\hline %inserts single line

\end{tabular}
%}
\label{tab:psd} % is used to refer this table in the text
\end{table}
%%%
%\section{Manuscript styles} \label{sec:style}
%\section{Observations and Data Reduction} \label{sec:two}
%\subsection{NuSTAR} \label{subsec:two-one}
%We analyze the six NuSTAR observations taken between February 16 and April 20, 2025 (see Table \ref{tab:log}).
%%%%
%Epoch 6, however, differs significantly from the earlier observations, as it does not exhibit any detectable QPO within the analyzed frequency range. One possible explanation is the substantial decrease in source flux during this final epoch, which may have reduced the signal-to-noise ratio to a level insufficient for QPO detection. Alternatively, it is possible that the QPO feature disappeared between the last two \textit{NuSTAR} observations, which are separated by 12 days. The rms varies only slightly among the epochs\footnote{Rms for each epoch computed in the $0.01-2$~Hz: 1) $ 27.0 \pm 0.2 \% $, 2) $ 28.2 \pm 0.2 \%$, 3) $ 26.9 \pm 0.2 \% $, 4) $ 27.6 \pm 0.3 \% $, 5) $ 24.5 \pm 0.3 \% $, 6) $ 26.0 \pm 0.6 \% $} ranging from $24.5\%$ in epoch 5 to $28.2\%$ in epoch 2.
% {\color{blue}[GM: here we could add some quantitative results, such as the total rms and the QPO freq]} 

Epoch 6 stands out from the earlier observations as it shows no detectable QPO within the analyzed frequency range. While this may reflect the disappearance of the QPO—a behavior commonly observed in black hole binaries during spectral evolution \citep{2016ASSL..440...61B}—an alternative explanation is that the QPO persists but with reduced amplitude or coherence, falling below the detection threshold due to the lower signal-to-noise ratio associated with the diminished source flux. Notably, the integrated broad-band rms amplitude remains relatively stable across all epochs, ranging from $24.5\%$ to $28.2\%$ in the $0.01-2\,\mathrm{Hz}$ range (see Table~\ref{tab:psd}), suggesting that the overall variability power is largely preserved.
The right panel of Fig.\ref{fig:PDS} focuses on Epoch 5. As with the spectral analysis, we divided the light curve into dip and persistent intervals and computed the corresponding PDS by averaging over each segment. As shown, the PDS from the two intervals are consistent within uncertainties above $0.1\,\mathrm{Hz}$, with the QPO frequency remaining unchanged despite the flux variation. The main difference emerges at lower frequencies (a few $\times 10^{-2}\,\mathrm{Hz}$), where the dip intervals exhibit enhanced variability. This low-frequency excess reflects the flux transitions between the high (persistent) and low (dip) states, which are intrinsically included in the dip segments.

\section{Discussion} \label{sec:four}
We report results from the spectral and timing analysis of the NuSTAR data of the black hole candidate IGR~J17091-3624 during its most recent failed outburst. We followed the hard-state evolution of the source over the course of nine weeks as it steadily decreased in flux and returned to quiescence---failing to transition out of the hard state. %displayed by the source during each of the four epochs.

\subsection{On the broadband spectral evolution of the failed outburst}
\begin{figure}%[ht!]
\includegraphics[width=.50\textwidth, angle=0, trim={2cm 1cm 2cm 1cm}]{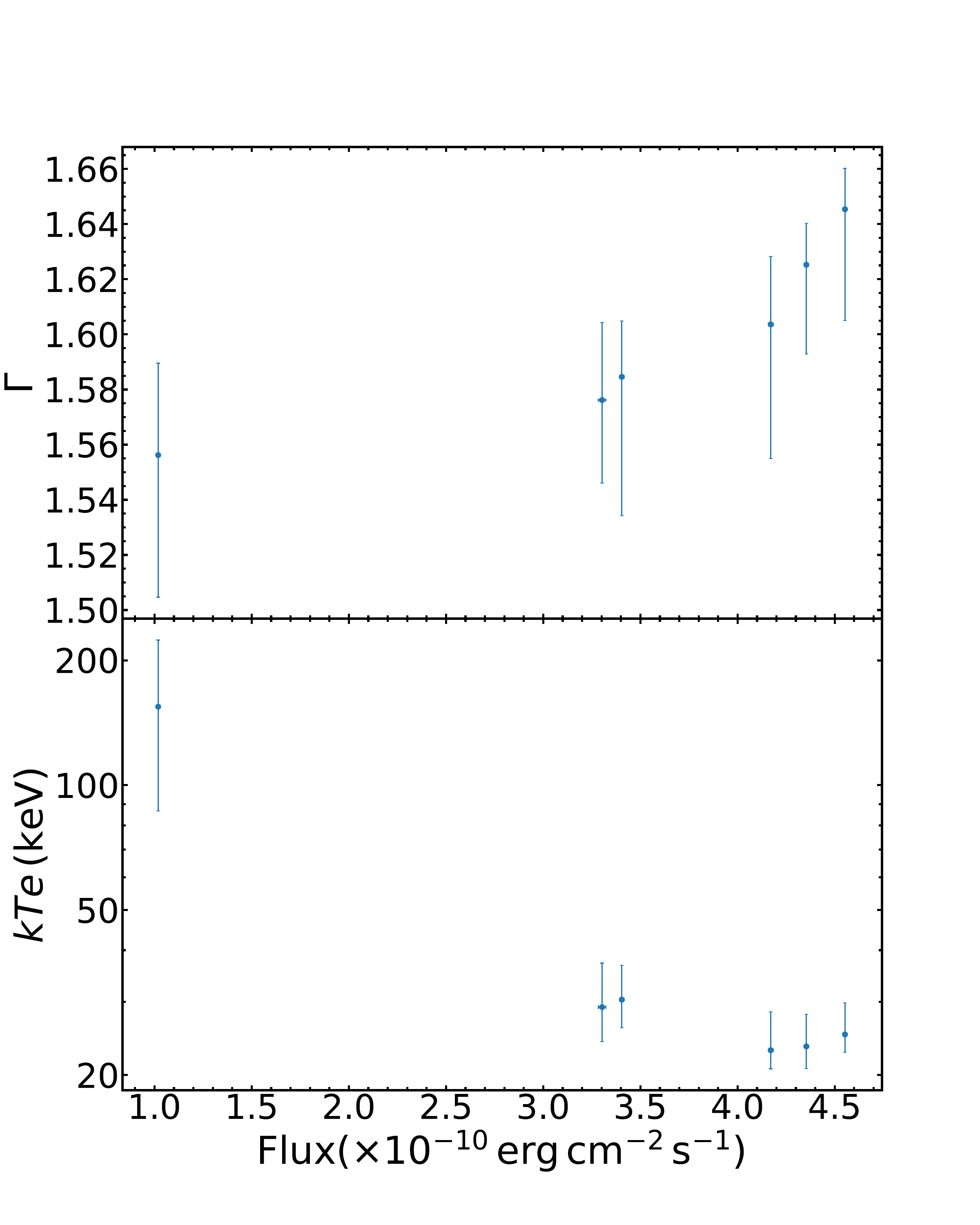}
\caption{Evolution of the photon index $\Gamma$ and the corona temperature $kT_{e}$ with the $2-10\,\mathrm{keV}$ flux for all six NuSTAR epochs. A straight line is fitted to data from each panel. The $2-10\,\mathrm{keV}$ flux decreased monotonically from Epoch 1 to Epoch 6. We note that the uncertainties on the fluxes are plotted but very small.} 
\label{fig:spectral_trends_1-6}
\end{figure}
During the course of the six NuSTAR observations, the spectrum remains hard, showing a decreasing trend in the value of $\Gamma$ with decreasing X-ray flux. Figure \ref{fig:epoch1-6_reflection_spec} shows that even at low accretion rates, relativistic reflection signatures, including the broad iron line and the Compton hump, are clearly seen in all epochs, indications of X-ray reprocessing in a low temperature ($\sim0.1\,\mathrm{keV}$), optically thick medium, most likely the accretion disk. Reflection spectroscopy consistently suggests strong relativistic reflection from a disk that is not significantly truncated, especially for Epochs 1-5. Additionally, the lamppost model returns a corona height between $\sim3-4\,\mathrm{R_{g}}$ for Epochs 1 to 5. By Epoch 6, the X-ray flux has dropped by a factor of $\sim3$ relative to Epoch 1, and although indications of disk reflection are still evident, they are not as strong compared to the first five epochs. In addition, the corona height has more than double in value. For each epoch, we computed the reflection strength, $R_{str}$, defined as the ratio of the estimated flux of \texttt{relxilllpCp} in the best fit model, to the unabsorbed flux, both estimated in the $0.1-100\,\mathrm{keV}$ band (see Table \ref{tab:mo_table}). For Epochs 1-5, the values of $R_{str}$ are comparable ($\sim20\%$). By Epoch 6, $R_{str}$ has dropped significantly to $\sim10\%$. All of these support the position that the disk properties did not evolve significantly over the course of the first five observations relative to Epoch 6. A qualitatively similar result was reported for GX~339-4 during its failed outburst in 2017 \citep{2019ApJ...885...48G}. By comparing their results with previous measurements of $R_{in}$ for the source, \citeauthor{2019ApJ...885...48G} posit that the inner accretion disk of GX~339-4 appears to be relatively close to the ISCO early in the outburst, reaching a few times $R_{\rm ISCO}$ at a luminosity level of about $1\%$ of its Eddington luminosity $L_{Edd}$. Our joint spectral analysis suggests that, over the course of the six observations, 
%while the disk temperature did not change appreciably, 
%the most notable changes are in the values of the photon index and the corona temperature (see Fig. \ref{fig:spectral_trends_1-6}). 
%If the apparent constancy of the disk properties is correct, 
%This would imply that 
the corona properties plausibly underwent significant changes from Epochs 1 through 6 while the disk reflection properties may not have changed appreciably through the first five epochs, plausibly due to a steady disk over this period. The trend in the estimated reflection strength, photon index and the coronal temperature tend to support this (see Table \ref{tab:mo_table}), although the uncertainties on $\Gamma$ are fairly large and the trend in $kT_{e}$ appears to be largely driven by the coronal temperature of Epoch 6 (e.g., see Fig. \ref{fig:spectral_trends_1-6}). Another caveat comes from the fact that NuSTAR is not very sensitive to changes in disk temperature, especially in the hard state.
%-with the caveat that NuSTAR is not very sensitive to changes in the disk properties especially in the hard state. 

In the hard state and before returning to quiescence, several observations have shown that the corona properties of BHXBs go through two regimes; ``softer when brighter'' and ``harder when brighter'' regimes---a ``V'' shape in the variation of $\Gamma$ with X-ray luminosity \citep[e.g.,][]{2008ApJ...682..212W, 2015MNRAS.447.1692Y, 2020ApJ...889L..18Y}. The turn-over typically occurs at bolometric luminosity, $L_{bol}\sim1\%\,L_{Edd}$. 
%(i.e. an X-ray luminosity of $\sim10^{36}\,\mathrm{ergs\,s^{-1}}$ for a $10\,M_{\odot}$ black hole). 
In quiescence, $\Gamma$ has been shown to saturate at $\sim2$ \citep[see e.g.,][]{2006ApJ...636..971C, 2013ApJ...773...59P}.
%an X-ray luminosity of $\sim3\times10^{36}\,\mathrm{ergs\,s^{-1}}$. 
Figure \ref{fig:spectral_trends_1-6} shows the variation of $\Gamma$ and $kT_{e}$ with the $2-10\,\mathrm{keV}$ flux for all six epochs for IGR~J17091-3624. The trend suggests that the source was still in the positive correlation ``softer when brighter'' branch over the course of the observations. This correlation is believed to be related to Compton cooling in the corona such that as the overall X-ray luminosity drops, the number of photons available to cool the corona via inverse-Compton scattering also drops, giving rise to a harder spectrum \citep[see e.g.,][]{2008PASJ...60..637M, 2009MNRAS.400.1603M}. 
\begin{figure}%[ht!]
\includegraphics[width=0.45\textwidth, angle=0, trim={2cm 1cm 2cm 1cm}]{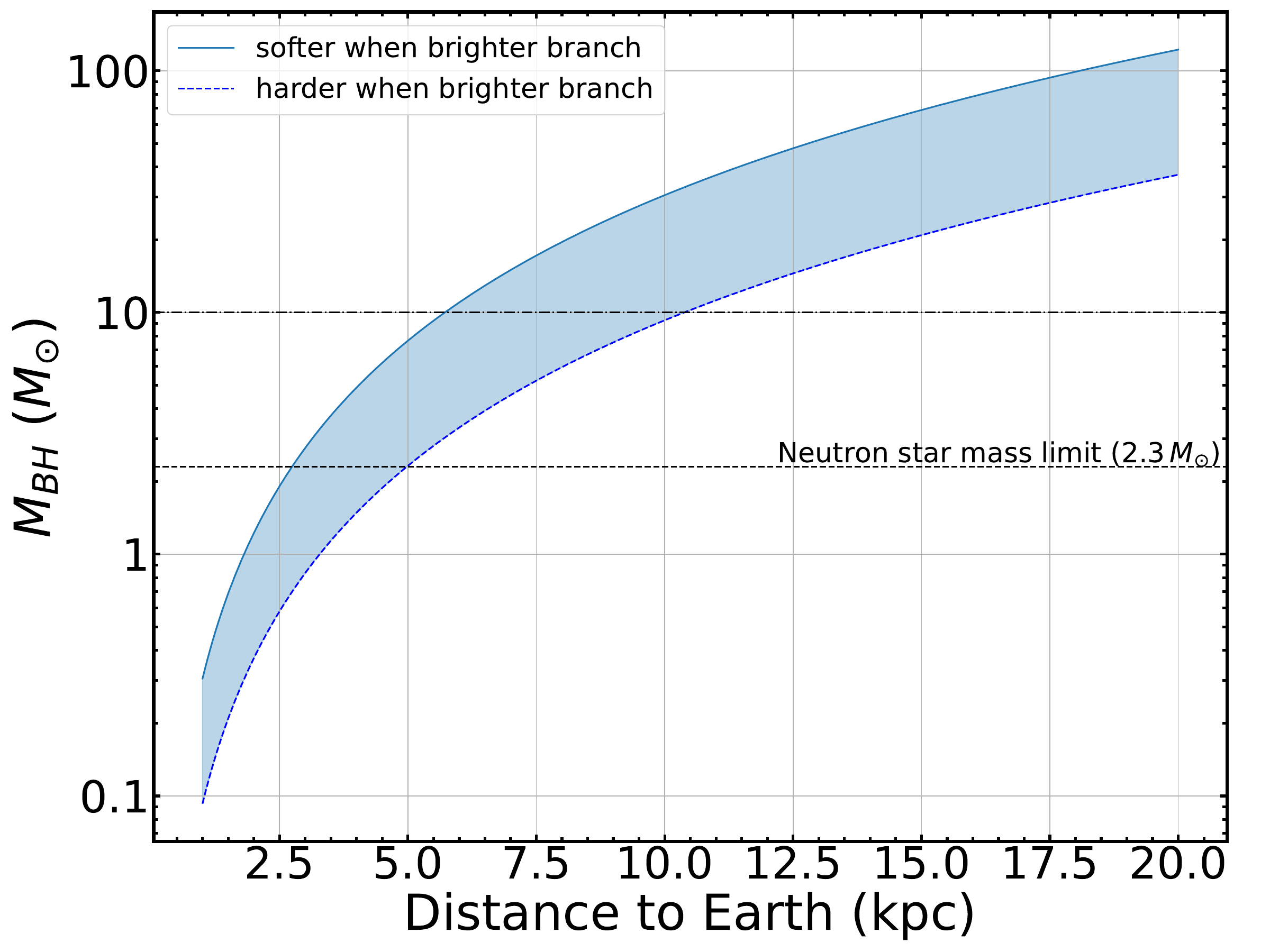}
\caption{Variation of black hole mass as a function of distance to the IGR~J17091-3624 system. The shaded band depicts the possible range of the distance to the system for a given mass and vice-versa. This is based on the $\Gamma-L_{X}/L_{Edd}$ correlation reported in \citet{2015MNRAS.447.1692Y} for BHXBs} 
\label{fig:mbh_distance_relation}
\end{figure}
Using data from thirteen BHXBs observed in quiescence as well as the hard and intermediate states, \citet{2015MNRAS.447.1692Y} obtained empirical relations between $\Gamma$ and the $2-10\,\mathrm{keV}$ luminosity ($L_X$) for the positive and negative correlation regimes (their Fig. 1). In the positive correlation (softer when brighter) branch, corresponding to $L_X/L_{Edd}\gtrsim0.001$, the relation is of the form  
\begin{equation}
\Gamma=(0.58\pm0.01)\mathrm{log}_{10}(L_{X}/L_{Edd})+(3.32\pm{0.02}).    
\label{eqn:one}
\end{equation}
In the negative correlation (harder when brighter) branch, corresponding to $10^{-6.5}\lesssim L_{X}/L_{Edd}\lesssim10^{-3}$, the relation has the form 
\begin{equation}
\Gamma=(-0.13\pm0.01)\mathrm{log}_{10}(L_{X}/L_{Edd})+(1.28\pm{0.02}).    
\label{eqn:two}
\end{equation}
Although Fig. \ref{fig:spectral_trends_1-6} suggests that IGR~J17091-3624 possibly remain in the positive correlation branch during the NuSTAR observations, the uncertainties are fairly large and the range of $\Gamma$ from all six epochs is significantly smaller than the range over which the correlation is derived from \citet{2015MNRAS.447.1692Y}. We therefore used equations \ref{eqn:one} and \ref{eqn:two} to estimate separately, $L_{X}/L_{Edd}$ using the average value of $\Gamma$ for all epochs (see Table \ref{tab:mo_table}).
%if we assume equation \ref{eqn:one} to be applicable, then using best fit parameter values from the joint fit to Epochs 1-6, the average value of $\Gamma$ and the $2-10\,\mathrm{keV}$ flux $F_{X}$ are $\sim1.59$ and $\sim3.47$, respectively. 
%Thus, the $2-10\,\mathrm{keV}$ Eddington-scaled luminosity of the source will be $\mathrm{log}_{10}(L_{X}/L_{Edd})\sim-2.99$  
For distance $d$ to the system, $L_{X}$ is related to the observed flux $F_{X}$ by the equation 
%\begin{equation}
    $L_{X}=4\pi d^{2}F_{X}$.
%\label{eqn:three}
%\end{equation}

With these simplifications, using the mean $2-10\,\mathrm{keV}$ flux for all epochs, we compute estimates for a range of possible masses for IGR~J17091-3624 as a function of the distance to the system. This is shown in Fig. \ref{fig:mbh_distance_relation}. The curves are plotted for both the positive and negative correlation regimes. As the figure illustrates, the positive correlation regime suggests that distances beyond $\sim10\,\mathrm{kpc}$ can probably be ruled out for IGR~J17091–3624, as they would imply a black hole mass exceeding $30\,M_{\odot}$---significantly above the observed upper range measured for low-mass BHXBs based on multi-wavelength studies. However, a larger distance remains plausible if IGR~J17091–3624 lies in the negative correlation regime. For a black hole mass of $10\,M_{\odot}$, shown by the horizontal dash-dot line in the figure, the predicted distance to the system is $\sim6-11\,\mathrm{kpc}$. This largely agrees with the lower bound predicted by \citet{2011A&A...533L...4R}. Figure \ref{fig:mbh_distance_relation} further suggests that the distance to IGR~J17091-3624 is unlikely to be less than $\sim3\,\mathrm{kpc}$ if it is to host a black hole rather than a neutron star. For the neutron star mass limit of $2.3\,M_{\odot}$, the predicted distance is $\sim3-5\,\mathrm{kpc}$. While this is simplistic at best, it provides some constraints on the distance to the system for a given black hole mass and vice versa. 
%Also, if either the distance or the mass is known, the other can be estimated. 
One major caveat here is that the curve is sensitive to the value of $\Gamma$ which can be model-dependent. Furthermore, it is not known with certainty if IGR~J17091-3624 follows the relations in Equations \ref{eqn:one} and \ref{eqn:two}. 
%be we make estimates for the possible mass of the source and its distance away using the relation in equation \ref{eqn:one}.

\subsection{On the Origin of the Dips}
The Epoch 5 observation unambiguously reveals several dipping intervals in its light curve---features that have never been reported for IGR~J17091-3624 before. At its deepest, the flux dropped to about $15\%$ of its persistent value. As Fig. \ref{fig:epoch5_e-resolved_lc} shows, the dips are more prominent at low energies, with a result that the dip spectrum is significantly harder. All of these suggest that the flux change is not intrinsic to the X-ray source. It is more likely caused by the obscuration of the primary source. Epochs 4 and 5 are separated by only $\sim23\,\mathrm{hr}$ and the NuSTAR light curve of Epoch 4 does not show any evidence of dipping. Also, the count rate and the broadband spectra during Epoch 4 are consistent with those from the steady portion of Epoch 5. Thus, the only difference between both epochs appears to be caused by photoelectric absorption and Compton scattering attributable to an obscurer passing along the line-of-sight to the compact X-ray source during Epoch 5. Spectral analysis confirms that the dip spectrum can be satisfactorily accounted for by obscuration from a moderately ionized ($\mathrm{log}~[\xi^{xstar}/\mathrm{erg\,cm\,s^{-1}}] \sim 2$) absorber having $N_{\rm H}\sim2\times10^{23}\,\mathrm{cm^{-2}}$---about 20 times the line-of-sight $N_{\rm H}$ in the direction of the source. Light-curve flux dips consistent with obscuration have been reported for a handful of BHXBs \citep[e.g.,][]{1998ApJ...494..747T, 2024ApJ...977...26A}. 
%Because these sources tend to be highly inclined
%%%%
%\begin{figure*}%[ht!]
%\includegraphics[width=0.9\textwidth, angle=0, trim={1cm 0.5cm 3cm 1cm}]{orbit_flux.pdf}
%\caption{Variation of the $0.1-100\,\mathrm{keV}$ X-ray flux irradiating the companion star as a function of the system's orbital period P (\textit{left}) and d, its distance away (\textit{right}), for assumed total system mass in the range $3-10\,M_{\odot}$. For each color shade, the solid and the dashed lines denote curves for $3\,M_{\odot}$ and $10\,M_{\odot}$ respectively. The dotted horizontal lines, at $5\times10^{11}\,\mathrm{ergs\,cm^{-2}\,s^{-1}}$, represents our chosen threshold beyond which the effect of ablation on the companion star is considered significant.} 
%\label{fig:dip_orbital-par}
%\end{figure*}
\begin{figure}%[ht!]
\includegraphics[width=0.45\textwidth, angle=0, trim={2cm 1cm 2cm 1cm}]{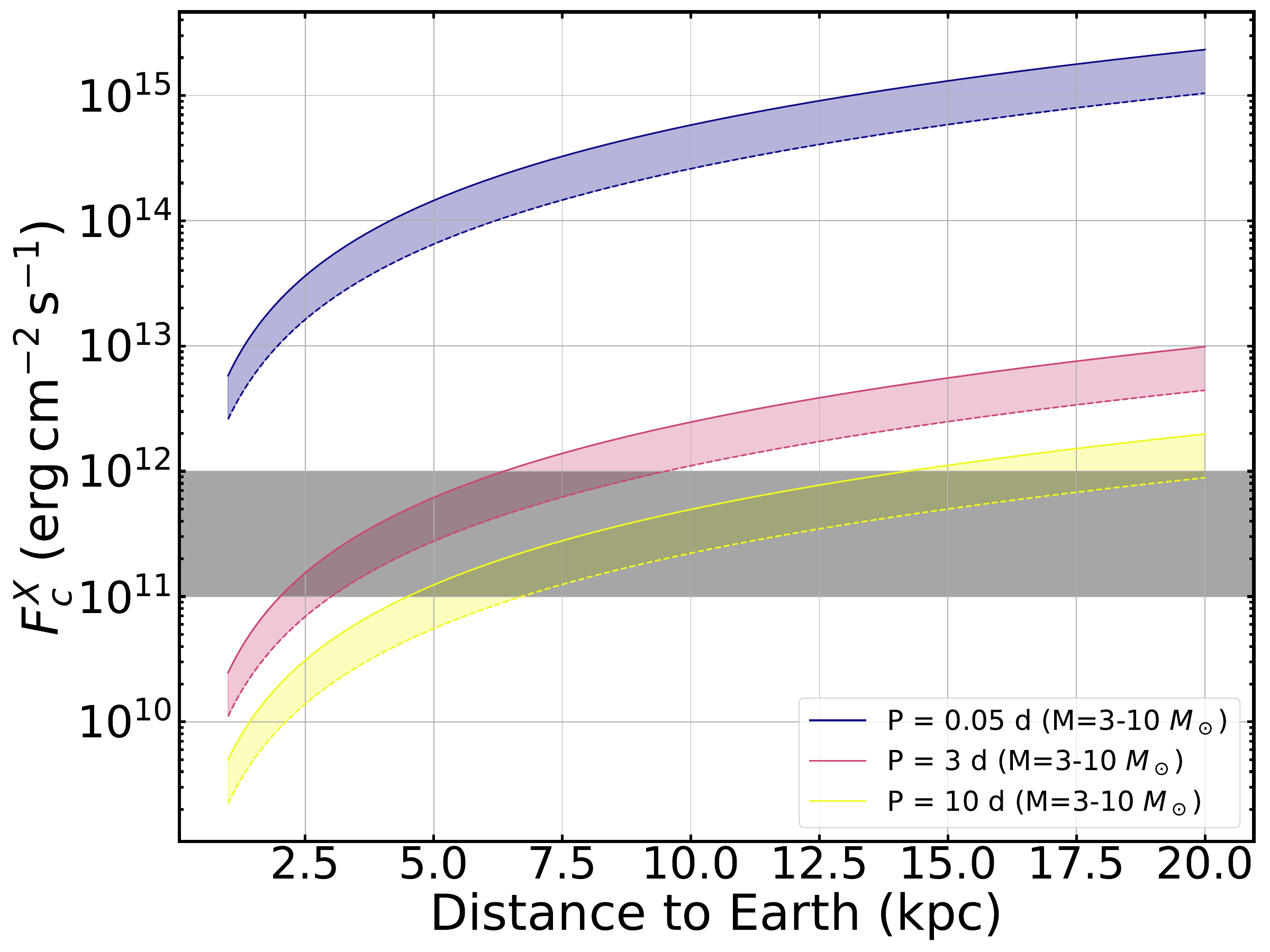}
\caption{Variation of the $0.1-100\,\mathrm{keV}$ X-ray flux irradiating the companion star as a function of the system's intrinsic X-ray luminosity, quantified by the distance to the system. Curves are plotted for three orbital period values for total system mass $M$ in the $3-10\,M_{\odot}$ range. For each color shade, the solid and the dashed lines denote curves for $3\,M_{\odot}$ and $10\,M_{\odot}$ respectively. The horizontal gray shade, at $10^{11}-10^{12}\,\mathrm{ergs\,cm^{-2}\,s^{-1}}$, represents the flux range where the effect of ablation on the companion star is considered significant as estimated by \citep{1991Natur.350..136P}.} 
\label{fig:dip_orbital-par}
\end{figure}
Because low-mass X-ray binaries that show dips but not eclipses typically have inclinations in the range $60\degree - 75\degree$, the obscurers are generally believed to be close to the disk plane \citep{1987A&A...178..137F}. 
%In high-mass X-ray binaries, the absorbing material is generally associated with clouds or blobs in the stellar wind of the companion star \citep[e.g.,][]{2000MNRAS.311..861B}. 
For IGR~J17091-3624, the spectral hardening during dips also points to this possibility, for which an absorber close to the disk plane absorbs the soft disk photons more effectively. A possible scenario is one where the stream of material from the secondary companion is thicker than the scale height of the accretion disk---presumably in the region where the accretion flow from the companion star impacts the outer accretion disk of the black hole. This may cause a fraction of the stream to flow above and below the disk. When such a material intercepts the irradiating X-ray continuum, ionization instabilities can separate the material into cold dense clouds, responsible for the dips, in a hot inter-cloud medium \citep[see e.g.,][]{1987A&A...178..137F}.
 
%%
%\begin{figure}%[ht!]
%\includegraphics[width=.50\textwidth, angle=0, trim={2cm 1cm 2cm 1cm}]{orbit_flux.pdf}
%\caption{Residuals from joint fit to data from Epochs 1-5 using \texttt{relxilllpCp} as base model. The data points are from FPMB} 
%\label{fig:dip_orbital-par}
%\end{figure}
%%
However, in a close binary system, when the black hole accretion rate is high, intense X-ray irradiation of the companion star by the black hole can drive a strong stellar wind even from a low-mass companion. The dips may thus be caused by X-ray absorption from clumps of such liberated partially ionized material directly coming along the line-of-sight \citep[e.g.,][]{2023MNRAS.520.3416K}. \citet{1991Natur.350..136P} estimated that in an interacting LMXB system, if the low-mass companion ($\lesssim1\,M_{\odot}$) is externally irradiated by X-ray flux in the range $\sim10^{11}-10^{12}\,\mathrm{ergs\,cm^{-2}\,s^{-1}}$, it will expand towards a new state of thermal equilibrium. Such external irradiation changes the star's effective surface boundary conditions, particularly by altering the degree of ionization of hydrogen at the bottom of the irradiated layer. Thus, potentially providing a new mechanism to drive mass transfer onto the compact object through a process similar to ablation.

In a close binary with black hole mass $M_{BH}$ and companion star mass $M_{CS}$, if the orbital period $P$ of the system is known, following Kepler's law the separation $a$ between the black hole and its companion is 
\begin{equation}
a=\left[\frac{G(M_{\rm{BH}}+M_{\rm{CS}})}{(2\pi)^{2}}P^{2}\right]^{1/3},    
\label{eqn:four}
\end{equation}
where $G$ is the gravitational constant. The X-ray flux irradiating the companion star $F_{c}^{X}$ is related to the observed X-ray flux $F_{o}^{X}\,(\approx2\times10^{-9}\,\mathrm{ergs\,cm^{-2}\,s^{-1}}$ for the persistent spectra of Epoch 5 in the $0.1-100\,\mathrm{keV}$ band) by the equation;
\begin{equation}
    F_{c}^{X}\approx F_{o}^{X}\left(\frac{d}{a}\right)^{2},
\label{eqn:five}    
\end{equation}
where $d$ is the distance to the system. The orbital parameters of IGR~J17091-3624 are not known and arguments have been made for the most extremes of parameters for the source based on its peculiar properties. The predicted estimates for the mass of the BH range from as low as $\sim3\,M_{\odot}$ to greater than $14\,M_{\odot}$. The distance to the system is also very uncertain, and the suggested orbital period of the system ranges from a few to tens of days \citep[see e.g.,][]{2011A&A...533L...4R, 2011ApJ...742L..17A, 2012ApJ...747L...4A, 2012MNRAS.422L..91W}. In Fig. \ref{fig:dip_orbital-par}, we show a plot of $F_{c}^{X}$ as a function of $d$, for a range of values for the total mass of the system between $3\,M_{\odot}$ and $10\,M_{\odot}$. For the X-ray flux measured at Earth, the distance to the system sets its luminosity or accretion rate. As one would expect, the plot shows that the further away the system is and the shorter the orbital period, the higher the chances of intense irradiation of the companion star by X-rays from the black hole. This can cause ablation of the outer layers of the companion star, potentially driving mass outflow via stellar wind even from a low-mass companion. As shown in the figure, the likelihood of ablation is significantly reduced for orbital periods longer than $10\,\mathrm{days}$, particularly if the system lies within $\sim15\,\mathrm{kpc}$. For orbital periods of $\sim3\,\mathrm{days}$, ablation becomes significant at distances beyond $\sim7\,\mathrm{kpc}$. At much shorter orbital periods of $\sim0.05\,\mathrm{days}$, substantial ablation is expected even if the system is no farther than $\sim2\,\mathrm{kpc}$. Analysis of the NICER observations covering the same dipping intervals seen with NuSTAR suggests an orbital period of $\sim3\,\mathrm{days}$, based on the inferred periodicity (private communication). If confirmed, this would strengthen the case for ablation as the origin of the dips and imply that the distance to the system is unlikely to be less than $\sim7\,\mathrm{kpc}$.

Relativistic reflection spectroscopy from previous outbursts of IGR~J17091–3624 consistently suggests a system inclination of approximately $30$–$40\degree$ \citep[e.g.,][]{2017ApJ...851..103X, 2024ApJ...963...14W}. However, the recurrent flux dips observed in our data, along with signatures of disk winds reported in earlier studies, are typically associated with high-inclination systems \citep[$\sim60$–$80\degree$;][]{2012MNRAS.422L..11P, 2024ApJ...977...26A}. Supporting this, IXPE observations overlapping with NuSTAR Epochs 4 and 5 revealed a high polarization degree, which \citet{2025MNRAS.tmp..846E} interpreted as evidence for a high inclination and/or substantial scattering in an optically thick disk wind or a mildly relativistic corona. Our own spectral modeling also yields an inclination of $\sim40\degree$, adding to the overall picture of a potentially complex geometry.
One possibility is that the inner disk (aligned with the spin axis) may be misaligned with respect to the outer disk (aligned with the binary's orbital plane) in the system \citep[e.g.,][]{2019ApJ...882..179C, 2021MNRAS.507..983L}. In this scenario, an observer viewing the system close to the outer disk plane may detect both signatures of disk winds and obscuration happening at the outer edges of the disk, whereas disk reflection coming from the inner disk would be consistent with lower inclination. Notable is the fact that \citet{2024ApJ...969...40D} found, from relativistic reflection modeling, that different spectra for a source can sometimes return conflicting inclination values. The authors suggested that this may be due to variable disk winds obscuring the blue wing of the relativistic iron K emission line. The fact that reflection modeling consistently yields low inclination values for IGR~J17091-3624 may rule out such a possibility for this source.
%A number of explanations have been put forward for how the orientation of a low-mass X-ray binaries to our line of sight determines the visibility dips which are presumable common in these systems. One possibility is a scenario where the
On the other hand, it is possible that the orbital inclination of the system is in a unique range between the population of face-on and edge-on systems such that the normal activities of the disk occasionally allow material at the outer edges to cross the line-of-sight \citep[e.g.,][]{2016MNRAS.461.3847G}.
%\subsection{Reconciling inclination with flux dips and disk winds in IGR~J17091-3624}

\subsection{On the timing evolution of the failed outburst}
Our timing analysis reveals a stable QPO at $\sim0.21\,\mathrm{Hz}$ across the first five epochs of the 2025 outburst, in contrast to the evolving QPO behavior observed during the 2016 event \citep{2017ApJ...851..103X}. In that study, \citeauthor{2017ApJ...851..103X} analyzed three \textit{NuSTAR} observations obtained during the rising phase of the hard state, for which the first epochs from both campaigns are at comparable flux levels. Over a 7-day interval in 2016, with observations spaced 5 and 2 days apart, the QPO frequency increased from approximately $0.13\,\mathrm{Hz}$ to $0.33\,\mathrm{Hz}$. Secondary peaks were also prominently detected at $\sim2.3$ times the fundamental frequency in each epoch. By contrast, during the 2025 outburst, the QPO frequency remained remarkably stable at $\sim0.21\,\mathrm{Hz}$ over a 20-day period spanning Epochs 1–5, pointing to a markedly different temporal evolution in the source’s variability properties. It is notable that during the 7-day span of the 2016 observations, the NuSTAR count rate increased by $\sim50\%$, while during the 20-day period covering Epochs 1–5 of the 2025 outburst, the count rate decreased by only $\sim25\%$. This more modest flux evolution may help explain the observed stability of the QPO during the 2025 event. 

The QPO frequency remains unchanged between the persistent and dip intervals, indicating that the structure of the inner accretion flow remains stable during the dips. This frequency stability, despite significant flux variations, strongly suggests that the dips are not caused by intrinsic changes in the innermost regions of the accretion disk. Instead, the enhanced low-frequency variability observed during dips likely arises from obscuration or absorption by inhomogeneous material located farther out in the disk or in the disk atmosphere. These findings support a scenario in which the inner accretion geometry remains intact, while the dips result from external structures intermittently crossing our line-of-sight. The clear decoupling between QPO behavior and dip-induced flux changes reinforces a geometric or absorption-based origin for the dips, consistent with models involving partial covering by clumpy or warped outer disk material.

\section{Conclusion} \label{sec:five}
Since its first known outburst a few decades ago, IGR~J17091-3624 has gone into outburst close to a dozen times. The most recent 2025 outburst, which turned out to be a failed outburst, is the first time that the source is observed to show flux dips in its light curves consistent with photo-electric absorption from an external absorber. We analyze and report results from the six NuSTAR observations of the source while the outburst lasted. Our findings are summarized as follow.

\begin{itemize}
    \item Over the course of the observations, the source shows significant relativistic reflection signatures especially during the first five epochs. Relativistic reflection modeling suggests a disk that is plausibly close to the ISCO and/or a very low corona height if a lamp post geometry is assumed for the corona.
    \item The spectral evolution of the source is consistent with a significant change in the corona properties---the photon index and the corona temperature---over the course of the outburst.
    \item The absorber material responsible for the observed dips during Epoch 5 may be produced by the ablation of the outer layers of the companion star, due to intense X-ray irradiation from the compact object. It could also, potentially, be from the normal accretion flow stream, when the size of the accreted material is larger than the scale height of the outer accretion disk at the point of impact. If caused by ablation, this puts useful constraints on the orbital period of and the distance to the system for a given mass.
    \item Our spectral modeling indicates a moderately low inclination ($\sim40\,\degree$), consistent with earlier spectral studies of the source. This inclination appears at odds with the observed dipping behavior and evidence of disk winds—features commonly linked to high-inclination systems. A possible explanation is a misalignment between the inner and outer disks, perhaps due to precession or warping. Alternatively, this inclination range ($\sim40$–$60\degree$) may represent a unique transitional regime where changes in the outer disk structure intermittently bring obscuring material into our line-of-sight.
    \item A spectral absorption line at $\sim7\,\mathrm{keV}$ is detected. While similar features have been reported during intermediate/soft states, their rarity in the hard state—combined with the line energy’s proximity to the iron K edge—makes it difficult to confirm whether the feature is intrinsic. The line is weak in the individual epoch spectra and becomes noticeable only in a joint fit across all six epochs. It is worth noting that at $6\,\mathrm{keV}$, NuSTAR’s effective area is significantly larger than XRISM’s \citep{2021SPIE11444E..22T}; thus, even if the line is real, the XRISM observation, partially overlapping with NuSTAR Epoch 2, may not detect it.
    \item Timing analysis reveals a stable $\sim0.21\,\mathrm{Hz}$ QPO across the first five epochs of the 2025 outburst, unlike the evolving QPO frequencies seen in 2016 for comparable flux levels. QPO was undetected in Epoch 6, possibly due to reduced flux. Low-frequency variability was enhanced during dips but the QPO frequency remained unchanged, indicating stability in the inner accretion flow.
\end{itemize}

%since figures and tables, see Section \ref{sec:floats}, will span the
%entire page, reducing the need for address float sizing.
 
%% Also note that the akcnowlodgment environment does not support long amounts of text. If you have a lot of people and institutions to acknowledge, do not use this command. Instead, create a new \section{Acknowledgments}.
\begin{acknowledgments}
The authors thank the anonymous referee for comments that improved the clarity of the manuscript. This work was supported under NASA Contract No. NNG08FD60C, and made use of data from the \textit{NuSTAR} mission, a project led by the California Institute of Technology, managed by the Jet Propulsion Laboratory, and funded by the National Aeronautics and Space Administration.
GM acknowledges financial support from the European Union’s Horizon Europe research and innovation program under the Marie Sk\l{}odowska-Curie grant agreement No. 101107057.
AI acknowledges support from the Royal Society.
J.B.C. is supported under 80GSFC21M0006. TD acknowledges support from the DFG research unit FOR~5195 (grant number WI 1860/20-1).
EN’s research is supported by an appointment to the NASA Postdoctoral Program at the NASA Goddard Space Flight Center, administered by Oak Ridge Associated Universities under contract with NASA.
MP acknowledges support from the JSPS Postdoctoral Fellowship for Research in Japan, grant number P24712, as well as the JSPS Grants-in-Aid for Scientific Research-KAKENHI, grant number J24KF0244.
\end{acknowledgments}


\vspace{5mm}
\facilities{\textit{NuSTAR}}
%\facilities{HST(STIS), Swift(XRT and UVOT), AAVSO, CTIO:1.3m,

\software{%astropy \citep{2013A&A...558A..33A,2018AJ....156..123A},  
          XSPEC \citep{1996ASPC..101...17A}, 
          XSTAR \citep{2001ApJS..133..221K},
          relxill \citep{2014MNRAS.444L.100D, 2014ApJ...782...76G}          
          }

\bibliography{bibtex}{}
\bibliographystyle{aasjournal}



\end{document}